\documentclass{aa}

\usepackage{graphicx}
\usepackage{natbib}
\usepackage{url}
\usepackage{listings}
\usepackage{color}
\usepackage[varg]{txfonts}
\bibpunct{(}{)}{;}{a}{}{,} 

\begin{document}

\newcommand{\Lya}{Ly$\alpha$}
\newcommand{\OI}{\mbox{[O\,\sc {i}]}}
\newcommand{\OII}{\mbox{[O\,\sc {ii}]}}
\newcommand{\OIII}{\mbox{[O\,\sc {iii}]}}
\newcommand{\CIII}{\mbox{C\,\sc {iii}]}}
\newcommand{\Ha}{H$\alpha$}
\newcommand{\Hd}{H$\delta$}
\newcommand{\Hb}{H$\beta$}
\newcommand{\Hg}{H$\gamma$}

\definecolor{light-gray}{gray}{0.95}
\defcitealias{herenz17}{H17}

\title{The MUSE-Wide Survey: Survey Description and First Data Release
\thanks{Based on observations carried out at the European Organisation
for Astronomical Research in the Southern Hemisphere under ESO
programs 094.A-0205, 095.A-0240, 096.A-0090, 097.A-0160 and 098.A-0017}}

\titlerunning{The MUSE-Wide Survey}
\authorrunning{Urrutia et al.}

\author{T.~Urrutia\inst{\ref{inst1}} \and
  L.~Wisotzki\inst{\ref{inst1}} \and J.~Kerutt\inst{\ref{inst1}} 
  \and K.~B.~Schmidt\inst{\ref{inst1}} \and E.~C.~Herenz\inst{\ref{inst2}}
  \and  J.~Klar\inst{\ref{inst1}} \and R.~Saust\inst{\ref{inst1}}
  \and M.~Werhahn\inst{\ref{inst1}} \and C.~Diener\inst{\ref{inst3}}  
  \and J.~Caruana\inst{\ref{inst4},\ref{inst5}} \and D.~Krajnovi\'{c}\inst{\ref{inst1}}
  \and R.~Bacon\inst{\ref{inst6}} \and L.~Boogaard\inst{\ref{inst7}}
  \and J.~Brinchmann\inst{\ref{inst7},\ref{inst8}} \and H.~Enke\inst{\ref{inst1}} 
  \and M.~Maseda\inst{\ref{inst7}} \and T.~Nanayakkara\inst{\ref{inst7}} 
  \and J.~Richard\inst{\ref{inst6}} \and M.~Steinmetz\inst{\ref{inst1}}
  \and P.~M.~Weilbacher\inst{\ref{inst1}}}

\institute{
Leibniz-Institut f\"ur Astrophysik, Potsdam (AIP), 
An der Sternwarte 16, 14482 Potsdam, Germany
\email{turrutia@aip.de}\label{inst1}
\and
Department of Astronomy, 
Stockholm University, 
AlbaNova University Centre, 
106 91 Stockholm, Sweden\label{inst2}
\and
Institute of Astronomy,
University of Cambridge,
Madingley Road,
Canbridge CB3 0HA, UK \label{inst3}
\and
Department of Physics, 
University of Malta, 
Msida MSD 2080, Malta\label{inst4}
\and
Institute of Space Sciences \& Astronomy, 
University of Malta, 
Msida MSD 2080, Malta\label{inst5}
\and
Univ. Lyon, 
ENS de Lyon, CNRS, 
Centre de Recherche Astrophysique 
de Lyon UMR5574,
69230 Saint-Genis-Laval, France\label{inst6}
\and
Leiden Observatory, 
Leiden University, 
PO Box 9513, 2300 RA Leiden,
The Netherlands\label{inst7}
\and
Instituto de Astrofísica e Ci\^encias do Espa\c co, 
Universidade do Porto, CAUP, 
Rua das Estrelas, 4150-762 
Porto, Portugal\label{inst8}
}

\date{Received / Accepted}

\abstract{We present the MUSE-Wide survey, a blind, 3D spectroscopic
  survey in the CANDELS/GOODS-S and CANDELS/COSMOS regions. The final
  survey will cover $100\times1$ arcmin$^2$ MUSE fields. Each
  MUSE-Wide pointing has a depth of 1 hour and hence targets more
  extreme and more luminous objects over 10 times the area of the
  MUSE-Deep fields \citep{muse-deep}. The legacy value of MUSE-Wide
  lies in providing ``spectroscopy of everything'' without photometric
  pre-selection.  We describe the data reduction, post-processing and PSF
  characterization of the first 44 CANDELS/GOODS-S MUSE-Wide
  pointings released with this publication. Using a 3D matched
  filtering approach we detect 1,602 emission line sources,
  including 479 Lyman-$\alpha$ (\Lya) emitting galaxies with redshifts
  $2.9 \lesssim z \lesssim 6.3$. We cross-match the emission line sources to
  existing photometric catalogs, finding almost complete agreement in
  redshifts (photometric and spectroscopic) and stellar masses for our
  low redshift ($z<1.5$) emitters. At high redshift, we only find
  $\sim$55\% matches to photometric catalogs. We encounter a higher
  outlier rate and a systematic offset of $\Delta$z$\simeq$0.2 when
  comparing our MUSE redshifts with photometric redshifts from the
  literature. Cross-matching the emission line sources with X-ray
  catalogs from the {\it Chandra} Deep Field South, we find 127
  matches, mostly in agreement with the literature redshifts,
  including 10 objects with no prior spectroscopic
  identification. Stacking X-ray images centered on our \Lya\,
  emitters yields no signal; the \Lya\, population is not dominated
  by even low luminosity AGN. Other cross-matches of our emission-line
  catalog to radio and submm data, yielded far lower numbers of
  matches, most of which already were covered by the X-ray catalog. A
  total of 9,205 photometrically selected objects from the CANDELS
  survey lie in the MUSE-Wide footprint, which we provide optimally
  extracted 1D spectra of. We are able to determine the spectroscopic
  redshift of 98\% of 772 photometrically selected galaxies brighter
  than 24th F775W magnitude. All the data in the first data release -
  datacubes, catalogs, extracted spectra, maps - are available on the
  website \protect\url{https://musewide.aip.de.}}   

\maketitle

\section{Introduction}

The first observations of the Hubble Deep Field \cite[HDF;][]{hdf} with
the {\it Hubble Space Telescope} ({\it HST}) proved to be an enormous step
for the field of observational cosmology, revealing thousands of
galaxies in a seemingly empty patch of sky. The blind nature of the
HDF revolutionized our view of galaxies; by not observing the nearby
galaxies we already knew, but staring at a dark, unknown portion of
the sky at high Galactic latitude, we were able to get a deep unbiased
look of an otherwise unremarkable part of the deep, distant Universe.

The success of this program prompted further observations of such deep
fields, the deepest being the Hubble Ultra Deep Field
\cite[HUDF;][]{hudf06} observed with {\it HST} at different
wavelengths from the UV to the Near-IR. Often a sort of
``wedding-cake'' approach is undertaken in such extragalactic surveys
- a very small area observed for long exposure times to reveal the
faintest and/or farthest objects, a medium area/exposure time
component to achieve larger number statistics without losing but the
most faintest galaxies and a shallow, large area component designed to
peer at low redshift or rare luminous objects.

Two of those blind extragalactic surveys in ``empty'' fields on which
we want to focus for the rest of this paper are: the GOODS-South and
the COSMOS survey: 

(a) The Great Observatories Origins Deep Survey \cite[GOODS;][]{goods}
is a deep multiwavelength blind survey with the {\it HST's} Advanced
  Camera for Surveys (ACS) instrument and Spitzer IRAC/MIPS
instruments. It spans roughly 320 arcmin$^2$ in two separate patches of
sky surrounding the ultra-deep {\it HST} observations and has a
limiting magnitude of $\approx$28 in the ACS passbands. It complements
the deepest X-ray observations in the sky (Chandra Deep Field South -
CDFS; \citealt{cdfs} and Chandra Deep Field North - CDFN;
\citealt{cdfn}) and was carried out in the early 2000s. Today there
exists a variety of long exposure observations from various facilities
of the GOODS-South region, ranging from hard X-rays
\citep{nustar-cdfs}, through the Far-Infrared \citep{herschel-cdfs},
Sub-mm \citep{alma-cdfs} and radio \citep{vla-cdfs}. In the center
third of the survey {\it HST} additionally carried out deep Wide Field
Camera 3 (WFC3) observations in the so-called CANDELS-DEEP survey
\citep{candels,candels2}. 

(b) The Cosmic Evolution Survey \cite[COSMOS;][]{cosmos} comprised of
640 {\it HST} orbits provides a somewhat shallower but larger area
than the GOODS survey, covering a 2 deg$^2$ area to reduce cosmic
sample variance. This field also has extensive multiwavelength
coverage from the X-rays \citep{chandra-cosmos} through the
Far-Infrared \citep{herschel-cosmos} to the radio \citep{vla-cosmos},
including over 30 bands in optical and near-IR data
\cite[e.g.][]{laigle}. Again, there is a strip covered by the WFC3
CANDELS survey. 

These {\it HST} deep fields have been instrumental in improving our
understanding of galaxy evolution, especially regarding the morphology
of galaxies across cosmic time. Many of the studies also sought to
bring insights from the local star formation ``main sequence''
\cite[SFMS, e.g.][]{brinchmann04,noeske07} to higher redshifts looking
for morphological differences \citep{wuyts11,bell12}, finding the
progenitors of present day massive galaxies \citep{barro13} or limits
and cosmic time evolution of the SFMS \citep{karim11}. Most of the
studies take advantage of the multiwavelength complements on the deep
{\it HST} data, e.g. to infer star-formation rates from the far-IR/radio
data or black hole accretion from the X-rays. 

However, a severe bottleneck to fully exploit the deep images is
presented by the difficulties of performing spectroscopic
follow-up. To some extent these difficulties have been alleviated by
the usage of photometric redshifts, but the simultaneous estimation of
redshifts, stellar population mix, dust extinction, and also nebular
emission line contributions leads to ambiguous results for at least a
significant fraction of galaxies. \citep{wilkins13,stark14}

There have been extensive spectroscopic campaigns in these fields. In
the GOODS-S region most of these identifications were done using the
power of ESO's VIMOS and FORS multi-object slit spectrographs on the VLT
\citep{lefevre05,vanzella08,balestra10}. For the COSMOS survey an
enormous investment in terms of spectroscopy was made through the
zCOSMOS survey, which used the VIMOS MOS spectrograph to gather more
than 20,000 galaxy spectra in the COSMOS area \citep{zcosmos}. The
currently ongoing Deep Extragalactic VIsible Legacy Survey (DEVILS)
seeks to increase the number of spectroscopic redshifts in GOODS-S,
COSMOS and other deep extragalactic fields to 60,000 \citep{devils}.
However, all of these spectroscopic campaigns required a photometric
pre-selection for the slit placement, be it some sort of magnitude
limit, a potentially interesting spectral energy distribution (SED) or
a non-optical counterpart. As such, multiple visits to the same field
are necessary to reach an acceptable completeness level, as slit
placements tend to overlap otherwise. In addition, the slit alignment
restriction ensures that it is very hard to reach the maximum optical
flux corresponding to a source, sometimes resulting in dramatic slit
losses. The {\it HST} grism mode of WFC3 \citep{dressel18} addresses
some of these concerns dispersing light from every source on the chip
without the need for preselection. However, the dispersed spectra
often overlap, necessitating various visits a different dispersing
angles to get a complete spectral census. The low spectral resolving
power (R $<$ 200) and different bandpasses in the UV and Near-IR make
{\it HST} grism spectra complementary to the science presented here. 

The large field of view of the Multi Unit Spectroscopic Explorer
\cite[MUSE;][]{muse} can alleviate several of these problems. MUSE is
a second generation Very Large Telescope (VLT) instrument for integral
field spectroscopy in the optical (4750 -- 9350 \AA). In its Wide
Field Mode, it has a $1'\times1'$ field of view with a spatial
sampling of 0.2\arcsec\, for a total of approximately 90,000 spectra taken in one
exposure. Its $\sim$2.5\, \AA\, resolution is suited to resolve the \OII\,
doublet throughout its whole wavelength range and produces a final
datacube with approximately $300\times300\times3680$ voxels (volume
pixels). By essentially covering the whole field of view continuously,
we are not restricted to a photometric pre-selection for
identification and classification of objects in the sky. In addition,
by possessing a full 3D view of the sky, we can select the relevant
voxels according to the shape of the galaxy and/or an interesting
wavelength range. Many techniques optimized for imaging (2D) and
spectroscopy (1D) can now be expanded to a 3D analysis (see, for
example, 3D crowded integral field spectroscopy, \citealt{pampelmuse}
or emission line detection in 3D cubes, \citealt{lsdcat}).    

The capabilities of MUSE in deep fields was demonstrated already
during commissioning by pointing MUSE for 27h at a 1 arcmin$^2$ region
in the Hubble Deep Field South \cite[HDFS;][]{hdfs}. This deep
integration provided nearly 200 redshifts in one go spectroscopically,
including 26 \Lya-emitting galaxies (LAEs) without an {\it HST}
counterpart. In addition, the 3D-nature of the MUSE instrument let us
study the morpho-kinematics of distant star-forming galaxies down to
stellar masses of $\sim 10^8\, \mathrm{M}_{\odot}$
\citep{contini16}. It also led to the discovery of  extended
\Lya-halos in the circumgalactic medium (CGM) of individual high
redshift galaxies \citep{wisotzki16}, the proper accounting of which
results in steeper \Lya\, luminosity functions \citep{drake17,herenz19}. 

In this paper we present the MUSE-Wide survey, a blind 3D
spectroscopic survey with MUSE of selected fields in the CANDELS-DEEP
and CANDELS-COSMOS regions. MUSE-Wide complements the MUSE-Deep survey
of the Hubble Ultra-Deep Field \citep{muse-deep}, sharing several of
the science goals but targeting a much larger area at a
correspondingly higher flux limit. MUSE-Wide furthermore provides
contiguous optical spectroscopic counterpart information to the many
multiwavelength surveys in this area. In a previous paper, we have
already presented a catalog of emission-line objects, based on a
subset of 24 fields of the MUSE-Wide data \cite[hereafter
\citetalias{herenz17}]{herenz17}. In this paper we describe the first
complete data release of this survey based on the first 44 fields of
MUSE-Wide. Besides the curated datacubes, the data release contains
identified and classified emission and continuum-selected galaxies;
the emission-line catalog contained in this data release supersedes
the \citetalias{herenz17} one. All the data and searchable catalogs
are available at \url{https://musewide.aip.de}. Throughout this paper
we adopt a flat Universe, $H_0 =
70\,\mathrm{km\,s}^{-1}\mathrm{Mpc}^{-1}$, $\Omega_{\Lambda} = 0.7$
cosmology.    

\section{Survey Description and Science Goals}

\subsection{Survey Description}

\begin{figure*}[ht!]
\centering
\resizebox{\hsize}{!}{
\includegraphics[height=7.4cm]{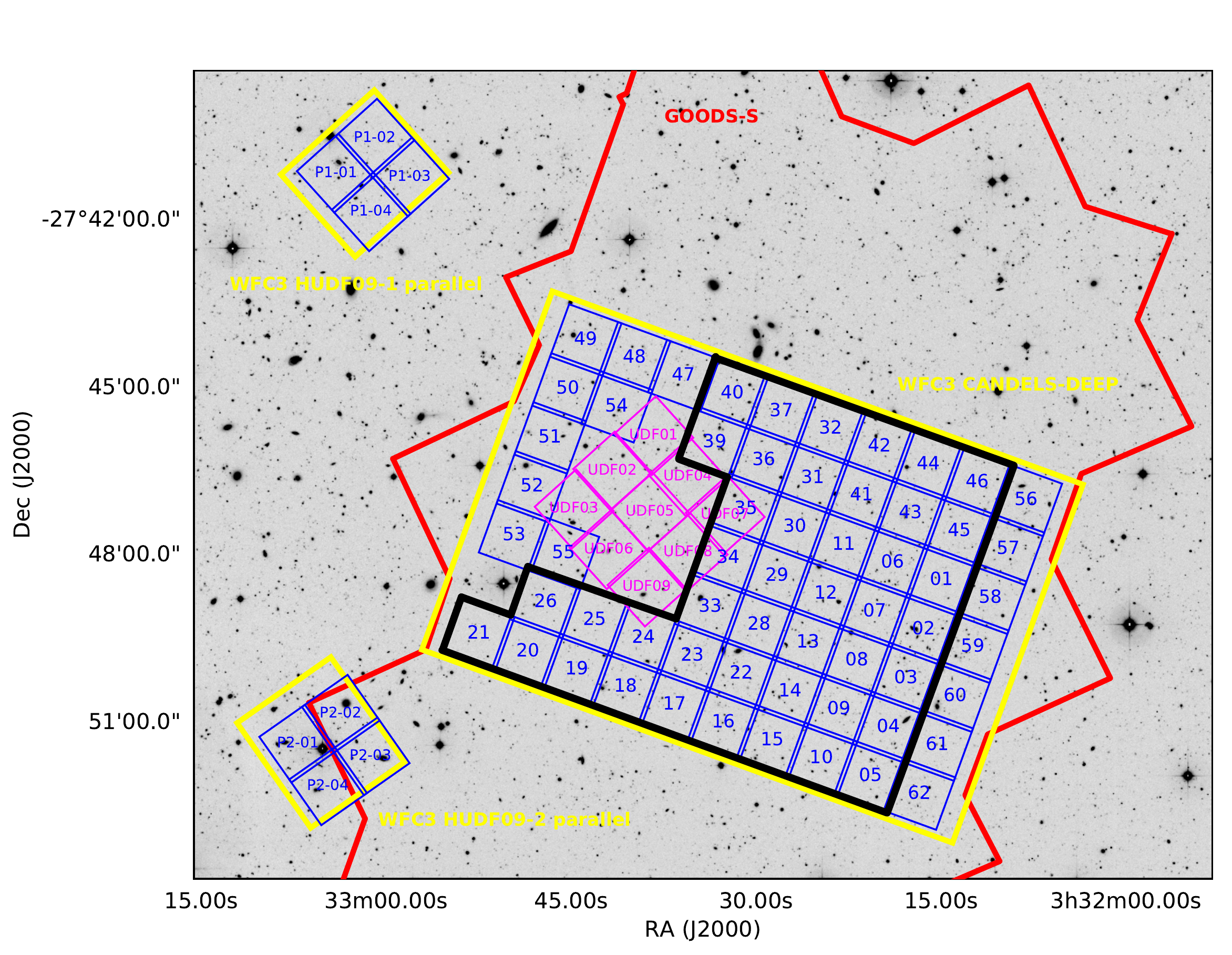}
\includegraphics[height=7.4cm]{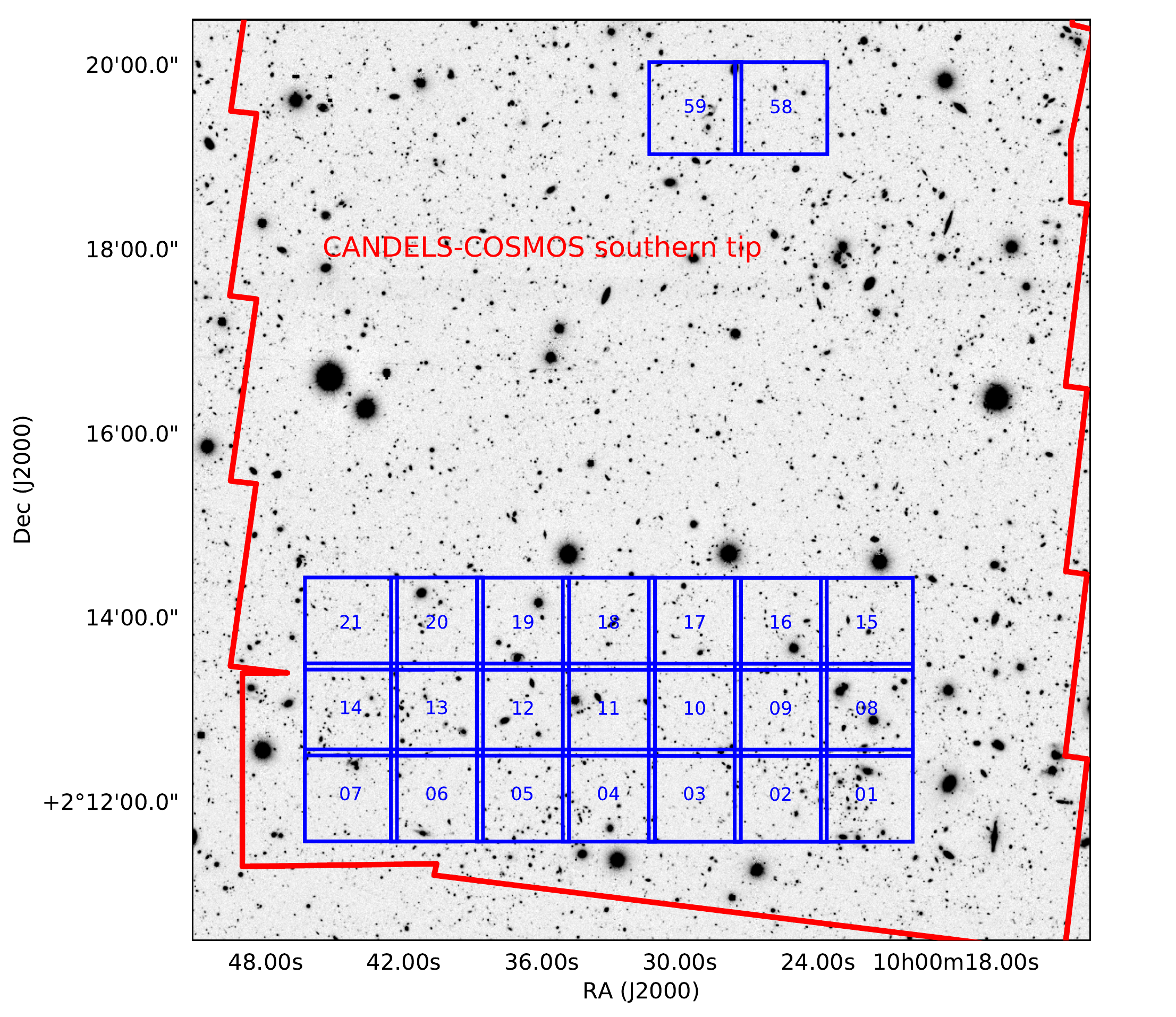}}
\caption{Layout of the 91 fields observed for the MUSE-Wide survey in
  blue. Left: The footprint in the Chandra Deep-Field South region
  overlaid on the V-Band image from the Garching-Bonn Deep Survey
  \cite[GaBoDS;][]{hildebrandt06}. Shown in red are the approximate
  contours of the GOODS-S ACS and in yellow of the CANDELS-DEEP and
  HUDF09 parallels WFC3 regions. The magenta regions represent the
  nine MUSE-Deep intermediate depth mosaic of the HUDF. The current
  data release encompasses the fields enclosed by the thick black
  line. Right: The footprint in the COSMOS region overlaid on the
  SUBARU-COSMOS $i$'-Band image \citep{subaru-cosmos} available from IRSA
  (\protect\url{http://irsa.ipac.caltech.edu/Missions/cosmos.html}). The
  red contours denote the southern tip of the deep {\it HST} exposures in
  the CANDELS-COSMOS region.} 
\label{musewide_layout}
\end{figure*}

MUSE-Wide is one of the Guaranteed Time Observing (GTO) programmes
executed by the MUSE consortium. It provides the ``wedding-cake''
observing approach often adopted in extragalactic surveys. MUSE-Deep
in the HUDF \citep{muse-deep} features a single 1 arcmin$^2$ area with
$\approx$31 hours observation time and a surrounding $3'\times3'$
mosaic covering the entire HUDF with $\approx$10 hours observation
time. MUSE-Wide covers $\sim10\times$ the area ($100\times1$
arcmin$^2$ fields with some overlap), but at only 1 hour observation
time. Yet, due to the excellent throughput of the MUSE instrument and
its use on on an 8m telescope, even 1 hour observations can reach
remarkably faint flux densities as we show below. 

MUSE-Wide mainly covers parts of the CDFS and COSMOS regions that were
previously mapped by {\it HST} in several bands to intermediate depths, by
GOODS-South in the optical \citep{goods} and by CANDELS in the near
infrared \citep{candels2}. The footprint of the individual MUSE-Wide
fields with respect to the {\it HST} coverage is shown in Figure
\ref{musewide_layout}.  We added 8 MUSE pointings in the so-called
HUDF parallel fields, specifically those parts that have deep Near-IR
imaging \citep[denoted as HUDF09-1 and HUDF09-2
in][]{parallels-wfc3}. Finally we also constructed and included
``shallow'' subsets of the MUSE-Deep data \citep{muse-deep} for the
purpose of checking our survey tools and classification strategy. This
way, MUSE-Wide comprises a total of 100 MUSE pointings. In our naming
scheme each field has a running number, preceeded by a region
identifier which can be either of the five: ``candels-cdfs'',
``candels-cosmos'', ``hudf09-1'', ``hudf09-2'', or ``udf''.  The
somewhat arbitrary numbering sequence of fields in the CDFS region
mainly reflects the order by which fields were added to the observing
queue over the semesters. Two fields (candels-cdfs-27 and -38) were
however removed from the list prior to observations, the former
because of the very bright star in the field, the latter because it
overlaps by more than 75\% with the udf-09 pointing of MUSE-Deep.  

Figure \ref{musewide_layout} shows the schematic mosaic tiling scheme
of all fields in MUSE-Wide. The combined footprint of the 44 fields in
data release 1 (DR1) is enclosed by the black line in the left side of
Figure \ref{musewide_layout}. Adjacent fields have a nominal overlap
of 4\arcsec as a buffer for telescope pointing and offset errors. The
candels-cdfs fields are oriented at a position angle (PA) of
$340\degr$ to match the CANDELS-Deep field layout. For similar
reasons, the hudf09-1 and hudf09-2 pointings were taken at a PA of
$42\degr$ and $35\degr$, respectively, candels-cosmos at PA of
$0\degr$, and the udf ``shallow'' again at $42\degr$. Here we present
and release the data for the first 44 candels-cdfs fields, i.e.\ with
field numbers 01--46. A future data release will encompass all 91+9
fields in both the COSMOS and CDFS areas. 
 
\subsection{Science Goals}\label{sciencecase}

As a ``blind survey of everything'' within the survey footprint and
sensitivity range, the design of MUSE-Wide clearly contains a strong
legacy aspect. However, the choice of fields and the observing
strategy were largely guided by our own scientific interests in these
data, which we briefly sketch out in the following. 

\subsubsection{A spectroscopic sample of 1000 Lyman-$\alpha$ emitting
  galaxies} 

Already more than 50 years ago, the \Lya\, emission line of hydrogen
was predicted to be a superb tracer for galaxy formation and evolution
studies in the high redshift universe
\citep{partridgepeebles67}. Meanwhile the study of \Lya-emitters
(LAEs) provides a route to identify low mass galaxies at high
redshifts that possibly constitute the progenitors of present-day
$L^\star$ galaxies such as the Milky Way \citep{gawiser07}. Most LAE
samples have so far been constructed from narrowband imaging
\cite[e.g.][]{HuMcMahon96,Rhoads00,Ouchi03,shibuya12,calymha,silverrush},
but significant efforts need to be spent on confirming LAE candidates
by spectroscopy. LAE samples have also been built from large
multi-object spectroscopic surveys \citep{stark10,cassata15}, but in
order to be efficient, such samples by construction rely on a very
stringent photometric preselection of high-$z$ candidates. The
all-in-one approach of using MUSE as a survey instrument obviates the
need of any pre-selection and follow-up spectroscopy. Given the
typical surface number density of about 10 LAEs detected per MUSE-Wide
field (\citetalias{herenz17}) we aim at building a sample of at least 1000
spectroscopically confirmed LAEs within $2.9 < z < 6.7$, all located
in fields with deep multi-wavelength data so that SEDs and physical
properties can be studied. Initial results from our first installment
of 24 fields include a measurement of the clustering properties of
LAEs \citep{diener17}, an estimate of the Ly$\alpha$ emitting fraction
among high-redshift galaxies \citep{caruana18} and a determination of
the \Lya\, luminosity function \citep{herenz19}. Already our first
sample of 237 LAEs constituted one of the largest existing sets of
high-$z$ \Lya\, spectra, especially when demanding a spectral
resolution good enough to study the line profiles in some detail
\citep{gronke17}.  

\subsubsection{Rare and extreme Ly$\alpha$ emitters}

A small fraction of LAEs appears to have \Lya\, rest frame equivalent
widths larger than the canonical limit of 200\,\AA\ for powering by
stellar processes from populations seen in galaxies today
\citep{kudritzki00,dawson04,gronwall07,kashikawa12,sobral15,hashimoto17a}. 
The reasons for such high equivalent widths are not well understood; a
possible explanation could be a higher ionizing continuum and a
consequently higher \Lya\, production rate at very low metallicities
\citep{raiter10}, and/or the enhancement of \Lya\, emission in very
recent bursts of star formation \citep{hashimoto17b}. The statistics of
these extreme objects is still quite poorly known, but they have been
posited to also be tracers for galaxies showing Lyman Continuum
leakage \citep{dijkstra16,marchi17}. While measuring rest-frame
equivalent widths higher than about 100~\AA\ is very difficult using
spectroscopy alone, the sensitivity can be greatly enhanced through
the combination with deep broadband continuum imaging, especially from
{\it HST}. Since most of the footprint of MUSE-Wide is within regions
covered by {\it HST} data of several orbits depth, MUSE-Wide provides
an exquisite dataset to search for LAEs with extremely high equivalent
widths. At $z=3$, associating a line flux of $2\times 10^{-17}$ erg
s$^{-1}$ cm$^{-2}$ (well above the 5$\sigma$ detection limit in most
of MUSE-Wide) with an object of continuum magnitude of 28 in the AB
system -- roughly the 5$\sigma$ limit in the ACS/F814W band of the
GOODS-S images -- would already imply a rest-frame equivalent width of
$\sim$400\,\AA, measurable with high significance. Even \Lya\,
equivalent widths of $>1000$\,\AA\ can still be measured confidently
by combining MUSE-Wide with {\it HST}, and we will be able to set
tight constraints on the occurence rate of such objects.  

Another still enigmatic category of objects are the so-called \Lya\,
``blobs''
\cite[LABs;][]{steidel00,bower04,nilsson06,weijmans10,erb11,matsuda11},
giant nebulae with often unclear associations to individual
galaxies. Given the recent MUSE discovery that essentially all LAEs
are also surrounded by extended \Lya-haloes
\citep{wisotzki16,leclercq17}, a clear-cut distinction between
``normal'' haloes and genuine LABs may be hard to draw. It is yet
unclear what the powering mechanism for LABs is; they could indeed be
powered by diverse sources of energy, such as extreme star formation,
cold accretion or AGN
\citep{prescott15,ao15,trebitsch16}. Nevertheless, the scales
associated with LABs make them different from the ubiquitous \Lya\,
haloes around low-mass star-forming galaxies. With estimated comoving
space densities between $10^{-6}$ and $10^{-4}$ per Mpc$^3$
\citep{yang11}, LABs are moderately rare objects. The total survey
volume of MUSE-Wide in \Lya\, amounts to roughly $10^6$ Mpc$^3$, large
enough that we expect to discover several new LABs, all of them with
already existing deep multiwavelength data. 

\subsection{Star-forming field low-mass galaxies at intermediate redshifts}

Studying galaxies of stellar masses around or below $\sim
10^8\,M_\odot$ is a difficult and expensive endeavour outside of the
local universe. Even when restricting this to star-forming galaxies
with strong emission lines, the extreme faintness of such objects
makes them hard to find and even harder to constrain their properties.
Yet such systems are of high astrophysical interest, as tracers of the
continued build-up of stellar mass several Gigayears after the peak of
the cosmic star formation history  \cite[e.g.,][]{behroozi13}, but
also as likely analogues to  low-mass galaxies at higher redshifts,
especially LAEs. In particular the so-called ``green peas''
\citep{cardamone09} have recently captured a lot of attention, not
least because of their possible relevance as leakers of Lyman
continuum radiation \citep{izotov16,izotov18}. Discovered by SDSS at
redshifts $z\la 0.3$, most known green peas are however too bright and
massive to be called genuine dwarfs. Shifting the known local
population of Blue Compact Dwarfs (BCDs) to $z\sim 0.5$ would result
in continuum magnitudes $V \ga 26$, too faint for nearly all recent
redshift surveys. 

Such objects are, on the other hand, easily detected in MUSE datacubes
from their conspicuous emission lines, as demonstrated by
\cite{paalvast18}, with $\sim 50$\% of the sample having stellar
masses below $3\times 10^{8}\,M_\odot$. Again, MUSE-Wide provides an
ideal hunting ground to find and characterise such systems, given the
broad spectral range of MUSE and the huge amount of complementary data
available. The wavelength range of MUSE complements blind {\it HST} grism
surveys, which probe for these low mass galaxies at different
redshifts and spectral resolutions \citep{wisp,maseda18}. For the
brightest galaxies of the sample, the 3D nature of MUSE data lets us
build two-dimensional maps of the gas kinematics \citep{guerou17}.

\begin{table*}
\caption{MUSE-Wide Observation Data}
\centering
\begin{tabular}{l c l c c} 
\hline\hline
Field  & Center Coordinates & UT Date Observed & Avg. Airmass & Avg. Seeing\tablefootmark{a} \\
 & RA Dec & (yyyy-mmm-dd) &  & [\arcsec] \\
\hline\hline
candels-cdfs-01 & 03:32:14.975 -27:48:29.36 & 2014-Oct-20 & 1.09 & 0.855 \\
candels-cdfs-02 & 03:32:16.416 -27:49:22.00 & 2014-Sep-20 & 1.02 & 1.045 \\
candels-cdfs-03 & 03:32:17.858 -27:50:14.63 & 2014-Nov-17 & 1.04 & 0.929 \\
candels-cdfs-04 & 03:32:19.301 -27:51:07.25 & 2014-Nov-17 & 1.19 & 0.763 \\
candels-cdfs-05 & 03:32:20.744 -27:51:59.88 & 2014-Nov-19 & 1.19 & 1.033 \\
candels-cdfs-06 & 03:32:18.941 -27:48:10.23 & 2014-Nov-18 & 1.39 & 0.844 \\
candels-cdfs-07 & 03:32:20.384 -27:49:02.86 & 2014-Nov-19 & 1.04 & 0.915 \\
candels-cdfs-08 & 03:32:21.826 -27:49:55.49 & 2014-Nov-19/20 & 1.49 & 0.996 \\
candels-cdfs-09 & 03:32:23.269 -27:50:48.12 & 2014-Nov-26 & 1.41 & 0.868 \\
candels-cdfs-10 & 03:32:24.713 -27:51:40.75 & 2014-Nov-27 & 1.28 & 0.899 \\
candels-cdfs-11 & 03:32:22.908 -27:47:51.10 & 2014-Nov-27 & 1.08 & 0.950 \\
candels-cdfs-12 & 03:32:24.350 -27:48:43.72 & 2014-Nov-27 & 1.11 & 1.023 \\
candels-cdfs-13 & 03:32:25.794 -27:49:36.35 & 2014-Nov-27 & 1.08 & 1.075 \\
candels-cdfs-14 & 03:32:27.237 -27:50:28.97 & 2014-Nov-28 & 1.10 & 0.883 \\
candels-cdfs-15 & 03:32:28.681 -27:51:21.60 & 2014-Dec-25 & 1.01 & 0.833 \\
candels-cdfs-16 & 03:32:32.649 -27:51:02.45 & 2014-Nov-28 & 1.02 & 0.825 \\
candels-cdfs-17 & 03:32:36.617 -27:50:43.28 & 2014-Dec-23 & 1.02 & 0.801 \\
candels-cdfs-18 & 03:32:40.583 -27:50:24.12 & 2014-Dec-21 & 1.02 & 0.885 \\
candels-cdfs-19 & 03:32:44.550 -27:50:04.94 & 2014-Dec-21 & 1.02 & 0.815 \\
candels-cdfs-20 & 03:32:48.517 -27:49:45.76 & 2014-Dec-23 & 1.11 & 0.820 \\
candels-cdfs-21 & 03:32:52.483 -27:49:26.57 & 2014-Dec-23 & 1.36 & 0.720 \\
candels-cdfs-22 & 03:32:31.205 -27:50:09.82 & 2014-Dec-22 & 1.02 & 0.790 \\
candels-cdfs-23 & 03:32:35.172 -27:49:50.66 & 2014-Dec-24 & 1.12 & 0.864\\
candels-cdfs-24 & 03:32:39.138 -27:49:31.50 & 2014-Dec-26 & 1.01 & 0.808 \\
candels-cdfs-25 & 03:32:43.105 -27:49:12.33 & 2015-Nov-05 & 1.11 & 0.788 \\
candels-cdfs-26 & 03:32:47.070 -27:48:53.14 & 2015-Oct-14 & 1.40 & 1.318 \\
candels-cdfs-28 & 03:32:29.761 -27:49:17.20 & 2015-Oct-11/12 & 1.24 & 0.963 \\
candels-cdfs-29 & 03:32:28.317 -27:48:24.58 & 2015-Aug-22 & 1.03 & 1.115 \\
candels-cdfs-30 & 03:32:26.874 -27:47:31.95 & 2015-Aug-21 & 1.15 & 1.091 \\
candels-cdfs-31 & 03:32:25.432 -27:46:39.32 & 2015-Aug-21 & 1.03 & 0.980 \\
candels-cdfs-32 & 03:32:23.989 -27:45:46.70 & 2015-Sep-10 & 1.03 & 1.075 \\
candels-cdfs-33 & 03:32:33.727 -27:48:58.05 & 2015-Sep-11 & 1.35 & 1.147 \\
candels-cdfs-34 & 03:32:32.283 -27:48:05.42 & 2015-Sep-11 & 1.11 & 1.120 \\
candels-cdfs-35 & 03:32:30.840 -27:47:12.80 & 2015-Sep-11 & 1.02 & 0.800 \\
candels-cdfs-36 & 03:32:29.397 -27:46:20.18 & 2015-Nov-10 & 1.13 & 0.940 \\
candels-cdfs-37 & 03:32:27.954 -27:45:27.56 & 2016-Feb-02 & 1.42 & 0.820 \\
candels-cdfs-39 & 03:32:33.361 -27:46:01.02 & 2016-Feb-03 & 1.15 & 0.813 \\
candels-cdfs-40 & 03:32:31.918 -27:45:08.40 & 2015-Aug-20/21 & 1.18 & 1.210 \\
candels-cdfs-41 & 03:32:21.465 -27:46:58.47 & 2015-Oct-12 & 1.14 & 0.985 \\
candels-cdfs-42 & 03:32:20.023 -27:46:05.84 & 2015-Oct-14 & 1.01 & 1.197 \\
candels-cdfs-43 & 03:32:17.500 -27:47:17.60 & 2015-Oct-14 & 1.14 & 1.155 \\
candels-cdfs-44 & 03:32:16.058 -27:46:24.97 & 2016-Mar-12 & 1.61 & 0.962 \\
candels-cdfs-45 & 03:32:13.533 -27:47:36.73 & 2016-Mar-13/14 & 1.50 & 0.904 \\
candels-cdfs-46 & 03:32:12.092 -27:46:44.10 & 2015-Oct-14/16 & 1.16 & 0.967 \\
\hline
\end{tabular}
\tablefoot{
\tablefoottext{a}{Gaussian in focal plane from Autoguider (best estimate).}}\label{observations}
\end{table*}

\section{MUSE Observations and Data Reduction}

\subsection{Observations}

The 44 candels-cdfs fields covered in this data release were observed
in 12 GTO runs from September 2014 to March 2016 (see Table
\ref{observations}). Most fields (80\%) were observed in dark time
with seeing just under or around 1.0\arcsec. A more detailed
description of the seeing properties is given in Section
\ref{psf-estimation} when discussing the Point Spread Function (PSF)
in the individual fields. 

Each MUSE-Wide pointing consists of 1h exposure time, split into $4
\times 900$s with 90\degr\, rotation in between and small fixed
dithers between the single exposures. The 4 exposures did not have to be
performed consecutively, they could be finished days later without any
consequence to the later combination of exposures, except the varying
observing conditions. The observations were carried out in nominal
mode, meaning each spectrum spans from 4750 -- 9350~\AA\, in
wavelength range, with the usual 0.2\arcsec$\times$0.2\arcsec spatial
and 1.25~\AA\, wavelength sampling, which is the default for
MUSE. Most pointings did not have bright enough stars for the slow
guiding system, hence we had to rely solely on the autoguider.

\subsection{Data Reduction}

All data were reduced with version 1.0 or with an early development
equivalent of the MUSE Data Reduction Software
\cite[DRS;][]{pipeline}. Although during the three years of
observations of all MUSE-Wide fields newer DRS versions and ancillary
software were released, we decided not to continuously update our
pipeline, but to reduce all cubes consistently in the same
manner. This ensured that quality differences are traceable solely to
observing conditions. 

The MUSE DRS operates in two stages. The first stage consists of
calibration recipes which work on the individual CCDs to determine or
remove the instrumental signatures of each IFU. At the end of this
stage a pixel table is created, which relates each of the 24 CCD x-y
positions and their flux values to a x-y-$\lambda$ position on a
datacube (Section \ref{prered}). In the second stage one or several pixel tables are
resampled onto a single datacube, usually with a 3D drizzling
algorithm. We performed the first stage processes with the usual
presets, but manipulated the pixel tables with our own routines before
combining them into individual datacubes (Section
\ref{slice-subtraction}). Furthermore, the combination of datacubes
was also performed with our own procedures and we added some
post-processing steps on the datacubes before arriving at the final
datacubes (Section \ref{selfcalzap} and \ref{effectivenoise}).

\subsubsection{Basic pre-reduction}\label{prered}

We produced the bias, flat, trace and dispersion master solutions from
the standard set of calibrations taken at the end of the night for
each MUSE-Wide observation. We applied these to the twilight skyflat
observations from which we then produced a twilight cube, describing the
unit-illumination correction.

After this, we applied the master solutions and the twilight flat to a
standard star observation taken either at the beginning or at the end of the
night. From these calibrated standard star exposures, we obtained the
system response curve and telluric correction for each night. This
response curve was further smoothed with a 30-order spline function to
get rid of small scale wiggles due to instrumental defects or sparse
sampling in the theoretical standard star spectrum. For each run we
compared the response curves to each other and if there were no
significant differences we used the response curve with the least
instrumental defects for flux calibration in that run. 

A new set of calibrations for the geometric and astrometric solution
of the MUSE instrument was obtained during each ovserving run. After
each science integration an additional illumination table (a short
lamp flat) was taken. This additional flat-field accounts for the
temperature variations in the flat-field, especially at the edges of
the IFU. Using all these calibration data, we removed the instrumental
signature from each CCD data and created the pixel tables.

\begin{figure}
\includegraphics[width=9.2cm]{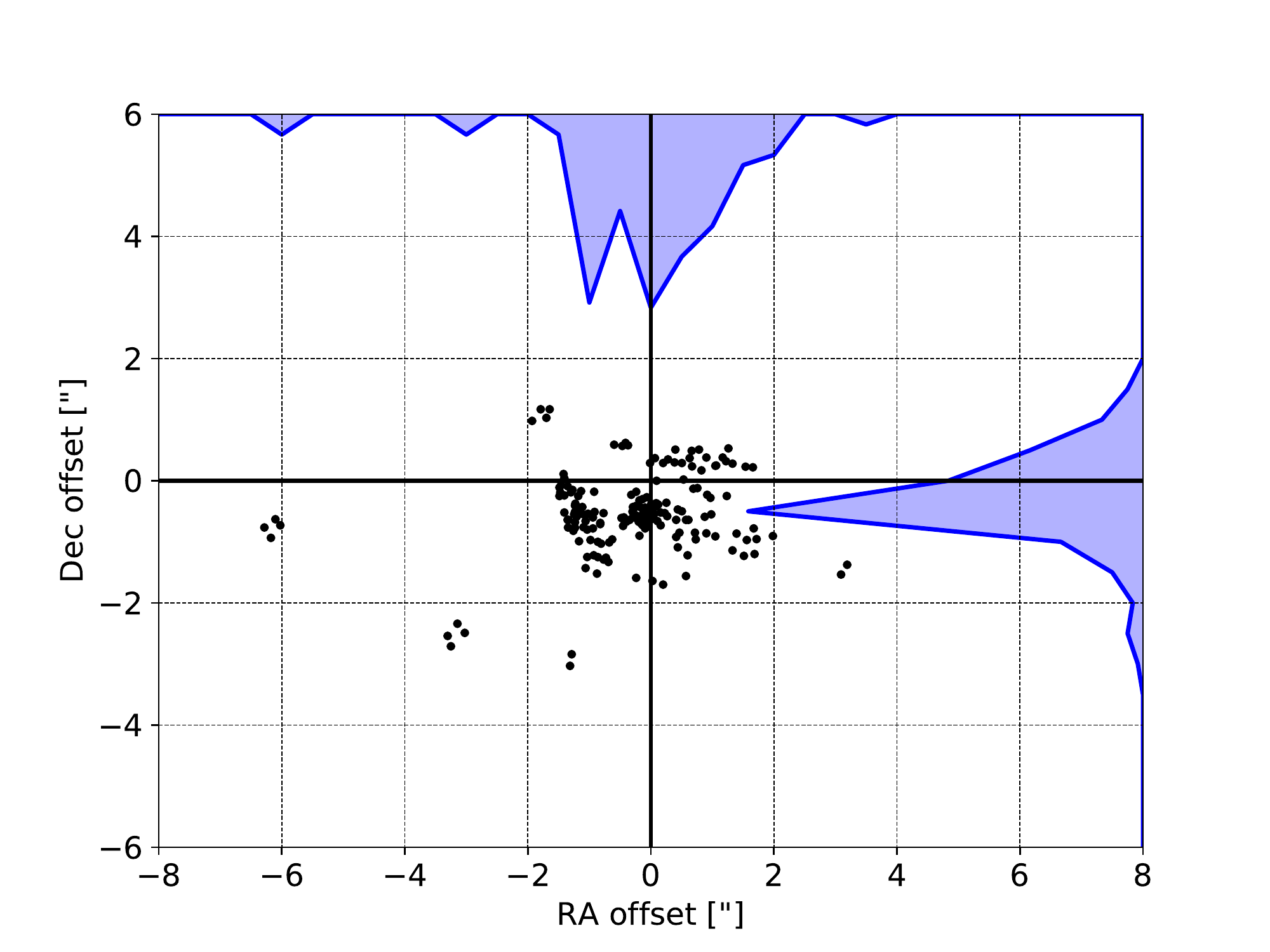}
\caption{Absolute RA/Dec offsets of the individual 15-minute MUSE
  exposures to the CANDELS WFC3 160W coordinates (shaded in blue are
  the offset distributions). Most offsets are around 1\arcsec. }\label{offsets}
\end{figure}

We created a first version of a datacube for each exposure using the
default values and the pipeline implemented sky subtraction. Using the
collapsed whitelight images from these cubes, we calculated the WCS
offsets by comparing a moderately bright star or 2 compact, moderately
bright galaxies to their WFC3-F160W CANDELS {\it HST} position (see Figure
\ref{offsets}). This needed to be done for each exposure as the
derotator wobble on MUSE introduced small offsets between
exposures. Most of the WCS offsets were around 1\arcsec\,and were
primarily due to night-to-night telescope misalignment. The
misalignment was especially pronounced in the beginning of November
2014 where the RA offset reached over 6\arcsec\, (see candels-cdfs-04 in
the exposure map in Figure \ref{large-expmap}). 

Instead of applying the offsets during the cube resampling, we applied
them to the reference World Coordinate System (WCS) in the pixel
tables manually, by subtracting them from the header values. This
later ensured that the four cubes had exactly the same sampling and
could be combined without the need for drizzling. We did create a
combined datacube using the WCS offsets, which was then our common
output grid on which all four exposures were be resampled. Before we
resampled, however, we applied our own sky subtraction and additional
flat-fielding to the pixel tables described below.  

\subsubsection{Slice-based Sky-Subtraction}\label{slice-subtraction}

The sky subtraction implemented by the MUSE pipeline
\citep{skysub-pipeline} is good to about the 2\% in areas outside of
significant sky lines. However, the remaining sky emission line
residuals are often significant and prevent us from reaching
background-limited sensitivity, especially for wavelengths redder than
wavelengths $\sim$7600~\AA. Instead we developed an alternative method
of sky-subtraction in MUSE data. Our approach works on the pixel table,
so that further post-processing, such as the self-calibration routine
described in Section \ref{selfcalzap} in the data reduction is possible. 

The main idea behind the method is the self-similarity of the line
spread-funtion (LSF) in the individual slices\footnote{When we refer
  to slices we mean the portion of light that is redirected by the
  MUSE image slicer and put through the pseudo-slits onto the CCD for
  each IFU. See Fig. 10 of the MUSE User Manual (version 8,
  \url{http://www.eso.org/sci/facilities/paranal/instruments/muse/doc/ESO-261650_MUSE_User_Manual_8.pdf})
  for a reference of the slice positions on the CCD.} of the CCD image
of each IFU. Since an emission line is sampled at just about 2 pixels
in width in the wavelength direction in the CCD plane, the tilt of the
slit and the curvature of the slices is crucial for the shape of the
LSF. A line that occurs in the leftmost slice of the CCD will have a
similar tilt and trace solution in all of the IFU CCDs. It is
therefore not necessary to model the LSF previously, each sky line
contribution is determined by an 24-IFU-average of the contribution
from each of the individual 48 slices. 

First we masked out the brightest 15\% and the dimmest 5\% of regions
in x-y datacube pixel coordinates (no WCS applied yet) to ensure
bright objects or instrumental defects did not interfere with a pure
sky spectrum. We created 48 ``sky slice spectra'' by taking the pixels
from all 24 IFUs on one slice (about 6.5 million) and averaging in
0.2~\AA\, bins, assuming that most of the slices contain empty sky and
by aggressive sigma-clipping (2.0$\sigma$) to get rid of emission
lines or cosmics.

In principle we now only had to subtract this ``sky slice spectrum''
by linearly interpolating it in wavelength and subtracting that
interpolation value from each pixel flux value in the pixel table
corresponding to that slice. Unfortunately the relative flux levels
between each IFUs due to small differences in the flat-field, were
significant enough to manifest themselves in the sky spectrum. These
IFU-to-IFU flux differences were typically less than 2.5\% in relative
value (though this also depends on the location of the slice relative
to the imaging edge), but were significant enough for the sky emission
lines to show significant IFU-to-IFU discrepancies when subtracting.   

\begin{figure*}[ht!]
\centering
\includegraphics[width=8.8cm]{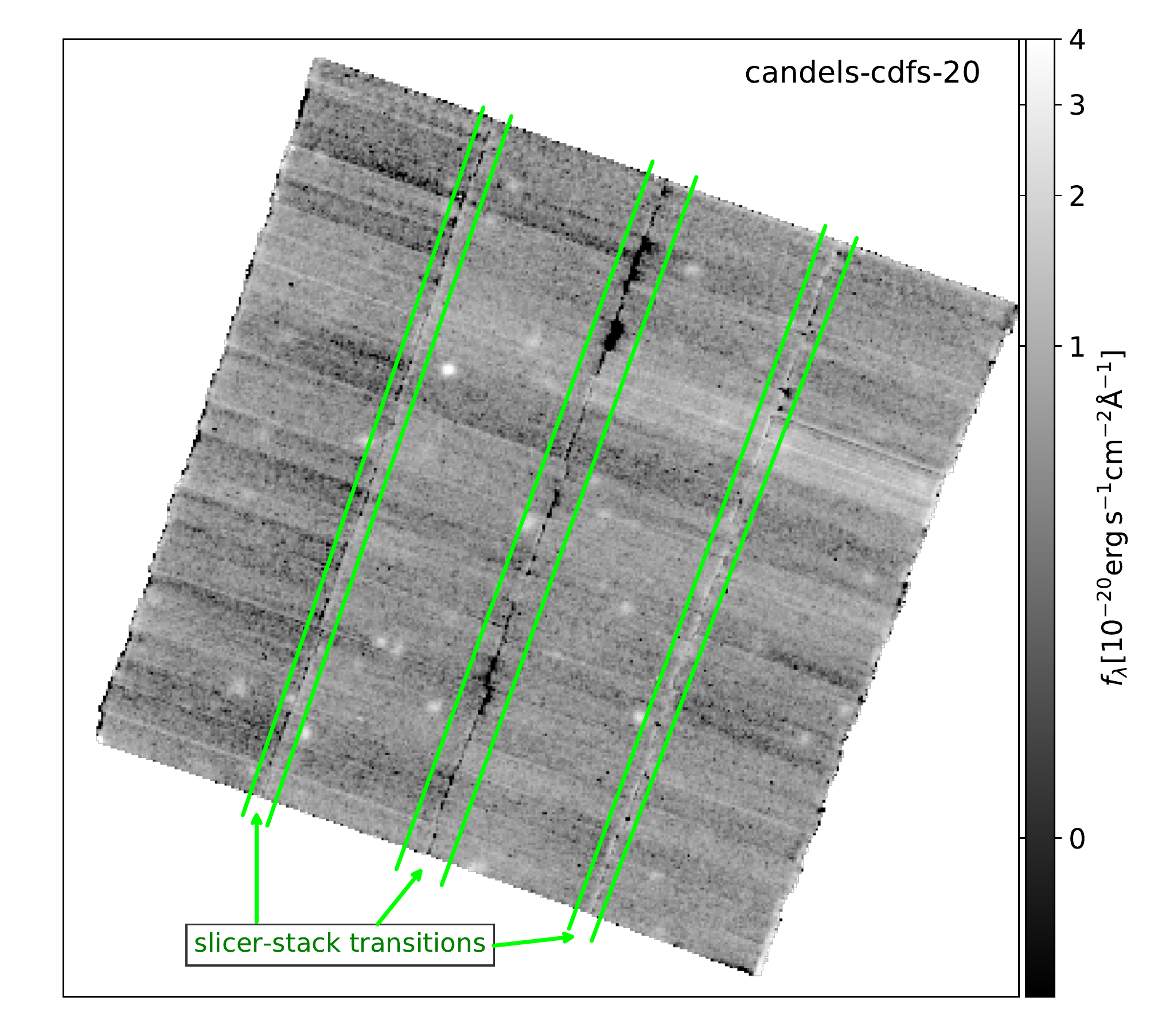}
\hspace*{0.2cm}
\includegraphics[width=8.8cm]{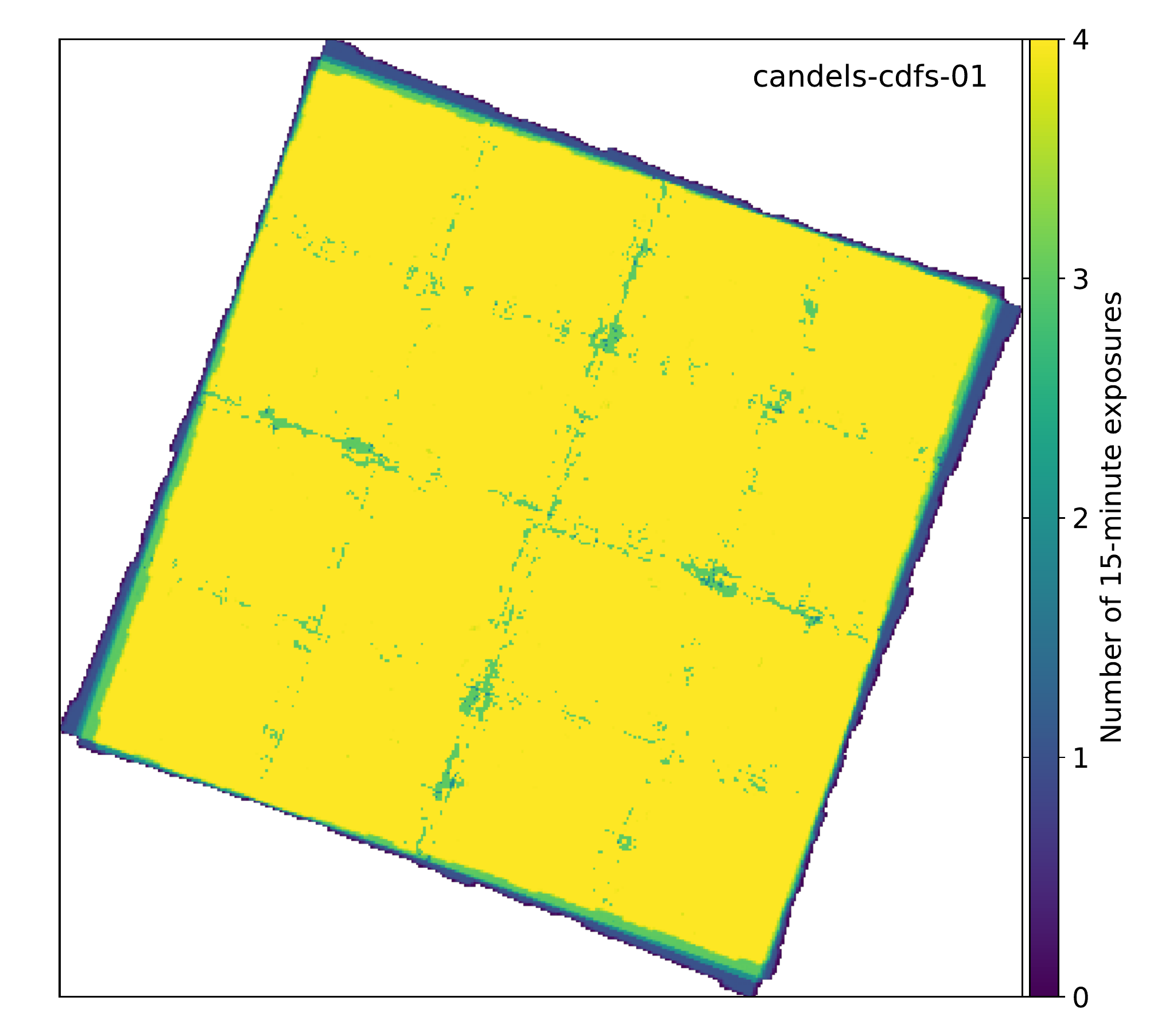}
\caption{(a) Left: Whitelight image of a single 15 minute exposure of
  a cube with only a few bright galaxies in the field (candels-cdfs-20),
  so that the contrast is enhanced. The dark regions in the slicer
  stack transitions immediately become visible. The regions in which
  the lowest 2\% voxels are masked are marked by the green straight
  lines. (b) Right: Example of a collapsed, combined 4x900s exposure
  cube (candels-cdfs-01) showing the different exposure times at the
  edges due to the trapezoidal shape of the MUSE field of view. The
  square pattern shows the regions masked in the slicer stack
  transitions.} 
\label{individual-expmap}
\end{figure*}

We determined the relative flux levels for each IFU by fitting a
Gaussian to 3 isolated sky emission lines (\OI\, at 5577.338~\AA\, and
6300.304~\AA\, and OH at 8943.395~\AA) across the spectrum for all 24
IFUs and all 48 slices in the pixel tables. While a Gaussian fit may
not describe the emission line perfectly, we were only interested in
the integrated flux, which to first order is conserved under changes
of the LSF. We employed all the pixel flux values that lay within
$\pm$4~\AA\ of the sky lines for the fit ($\approx550$ per sky
line). For each slice we then calculated the relative integrated flux
values of the sky lines for each IFU and used the median value of
these three to be the one to normalize the slice sky spectrum by for
each IFU. This normalized, interpolated spectrum was then subtracted
from each pixel flux value in the pixel tables. 

In fact, the initial assumption that each slice has the same LSF is
not strictly correct, hence there will still be residuals in the
subtracted sky-line regions. However, the residuals using this
slice-subtraction method are about 25\% in amplitude when compared to
v1.0 of the pipeline method. By working on the pixel tables, we could
apply further post-processing steps described in the next section,
including a second sky subtraction using principle component analysis
(PCA). In addition, the sky normalization applied before subtracting
can be interpreted as additional flat-field, ensuring greater
uniformity. 

\subsubsection{Further post-processing in the data reduction}\label{selfcalzap}

After we applied the slice-based sky-subtraction on the pixel tables,
we used the MPDAF \citep{mpdaf} self-calibration method to remove
systematic mean zero-flux level offsets between slices and IFUs. We
employed whitelight images of each slice-subtracted pixel tables
resampled to the common output grid as tracers for the mask applied in
the self-calibration. We note that this early version of the self-calibration
recipe \cite[comparable to the one used on the HDFS,][]{hdfs} still
showed the familiar striping pattern in collapsed MUSE images. After
we applied the self-calibration, we finally resampled the modified
pixel tables one last time to the common output grid; each volume
pixel in the datacube (voxel) had exactly the same 3D (RA, Dec,
$\lambda$) position.

\begin{figure*}[ht!]
\centering
\includegraphics[width=15cm]{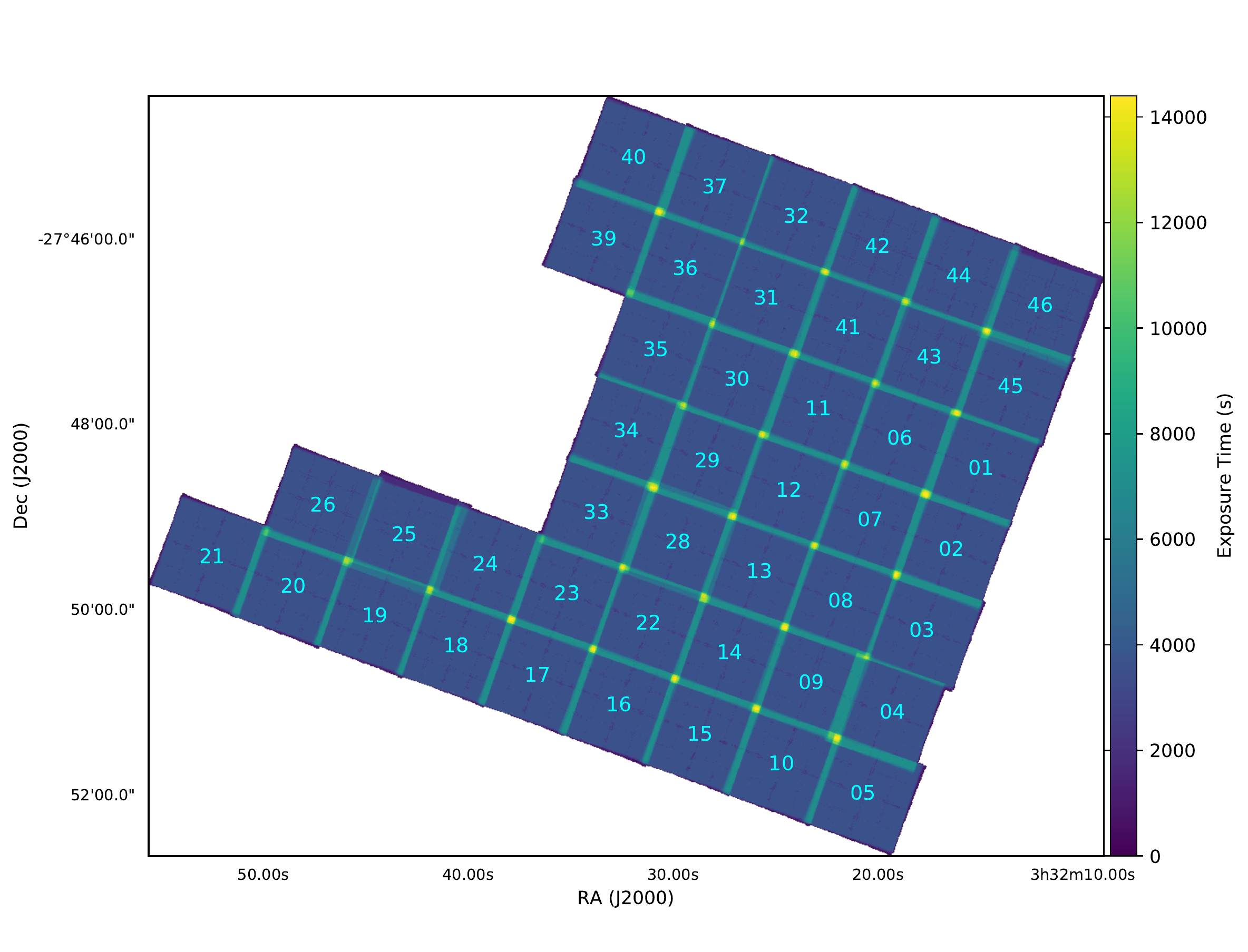}
\caption{Combined exposure maps of all 44 DR1 fields showing the
  coverage of the MUSE-Wide areas including the overlap regions which
  show up to 16 x 15 minute exposures.}\label{large-expmap}
\end{figure*}

Because of flux aberrations in the slicer stack transition areas, some
pixels at these transitions receive lower light levels, leading to
dark spots in the combined datacube. These aberrations are wavelength
dependent and are also seen to vary slowly over time. Since with the
MUSE-Wide observing strategy each point in the sky ended up in four
different IFUs and positions relative to the slicer stack, we dealt
with this phenomenon as a cosmetic defect by a simple masking
strategy. We masked out all the voxels that showed the lowest 2\% of
flux in the whitelight images and lay within the slicer stack
transitions regions enclosed by straight lines (see Figure
\ref{individual-expmap}a). When the 4 exposures were taken
sequentially, there was little zero-point offset between the
exposures, so the same straight-line regions could be used. Only when
the exposures had significant pixel shifts with respect to the common
output grid was there a need to set the masking regions manually.  

Before we combined the masked cubes, we performed a second sky
subtraction on the individual exposure datacubes to remove some
after-residuals due to the varying shape of the LSF. We used ZAP v1.0
\citep{zap}, a method taking advantage of the sky lines affecting all
the voxels of the whole cube equally.  

Finally we combined the four ``ZAPed'' individual 15-minute
datacubes. The flux cubes were averaged with a 3 sigma clip to exclude
any extreme outliers known to be prevalent at the edges. The variance
cubes were averaged (without sigma clipping) and divided by the square
of number of exposures, capturing the masking and the different
exposure levels at the edges. As explained below, the combined
variance cube was subsequently replaced by a self-calibrated
``effective variance'' cube corrected for resampling (see Section
\ref{effectivenoise}). 

We created a whitelight image by masked averaging over all the pixels
of the flux datacube in wavelength direction. Finally, we created an
exposure cube with values between 0 and 4 by summing up the individual
exposure cubes, which consisted simply of a zero if there is a NaN
value in the cube and 1 if there is not. The exposure cubes do not
only capture the edges and the slicer-stack transition masking, but
also masked out cosmic ray regions which only affect a small
wavelength range or sparse coverage at the wavelength boundaries of
4750~\AA\, and 9350~\AA.  

The two-stage sky subtraction procedure described above does a good
job of keeping the background level reasonably flat within each
spectral layer of a cube, especially across the instrumental stacks
and slices, but it does not ensure a zero expectation value for the
mean background level. We therefore added a postprocessing step to
estimate background correction values as a function of wavelength. We
first built a binary blank sky mask by thresholding the whitelight
image, followed by a sequence of binary filtering (erosion and
dilation), leaving typically $\sim$60\% of the field of view as
unmasked. We then calculate, separately for each spectral layer, the
mean of all blank sky pixels. Assuming that the expectation value of
the background correction in general varies slowly with wavelength, we
smooth the array of mean background values by a succession of spectral
median and Gaussian filters, which we then adopt as background
correction for most layers. An exception is made at wavelengths with
strong night sky emission lines and corresponding residuals, where we
use the monochromatic mean per layer without the spectral
filtering. The resulting background offset values are then subtracted
from the cube. These corrections tend to be small, in the range of
2$\times 10^{-20}$ erg\,s$^{-1}$cm$^{-2}$\AA$^{-1}$, but would add up
when integrating over large apertures.

All of these combinations were possible, because the individual cubes
had the same astrometry and astrometric zero-point and had been
resampled onto the same common output grid. The combined flux cube,
the combined variance cube, the summed exposure cube and the whitelight
image are all stored in a multi-extension FITS file. The
multi-extension datacube for one field takes about 5GB of disk
space. The cubes use air wavelengths (instead of vacuum) and are
corrected to heliocentric reference frame.

Figure \ref{large-expmap} shows a map of the collapsed exposure cubes
(so-called ``exposure whitelight images'') for the entire 44 fields
(created with IRAF \texttt{imcombine}). Of particular note is the
large WCS zero-point offset in candels-cdfs-04 and a large shift of
two 15-minute exposures taken 4 hours later in the night in
candels-cdfs-25. Using the entire exposure map we compute the solid
angle of the MUSE-Wide DR1 footprint with at least two 15 minute
exposures to be 39.5 arcmin$^2$.  

\subsubsection{Effective variances}\label{effectivenoise}

The voxel-by-voxel variances obtained by formal error propagation in
the MUSE pipeline systematically underestimate the true uncertainties
in the cube because of the resampling needed to construct the cube,
which shifts some of the power into covariances. Another disadvantage
of the formally calculated errors is that they are inherently noisy,
since they are based on actually measured count rates per voxel
instead of the corresponding expectation values. This second property
can lead to severe biases in the extraction of faint object spectra
when using weighting schemes based on voxel variances. In
\citetalias{herenz17} we tried to estimate the variances empirically by
measuring the median aperture flux in 100 random empty sky positions
in the MUSE flux cubes. We now describe an improved three-step
procedure to replace the variance cube provided by the pipeline with
empirically calibrated errors.  

(i) We measured, separately for each wavelength, the typical variance
between individual blank-sky voxels as $s^2 \approx [0.7413 \times
(q_{75}-q_{25})]^2$ where $q_{25}$ and $q_{75}$ are the 25\% and 75\%
quartiles of the distribution of voxel values at given spectral layer
and the factor 0.7413 rescales the quartile distance to an equivalent
Gaussian standard deviation. These variance estimates implicitly
include a contribution from small-scale systematics such as imperfect
flat fielding or sky subtraction. To distinguish them from other
approaches to quantify the uncertainties we denote these empirically
calibrated errors as \emph{effective noise}.  

(ii) To calibrate the bias arising from the resampling process we
created an artificial pixel table with normally distributed random
numbers with zero mean and unit variance and pushed it through the
pipeline resampling with the same setup as the observational data,
producing a cube containing only random numbers but including the
cross-talk between neighboring voxels, thus with formally propagated
voxel variances substantially smaller than unity. Assuming that the
resampling effects on the variances are the same for this random
numbers cube and for real datasets, we rescaled the empirical
voxel-by-voxel variances $s^2$ by a slightly wavelength-dependent
calibration factor $f_{\mathrm{rs},\lambda}$ (taken as the inverse of
the median variance in the random cube) to approximately account for
the losses due to resampling. We checked the correctness of this
calibration by comparing the resulting effective noise values to the
expected pure shot noise (without resampling) from the measured sky
brightness and the detector readout noise, finding good agreement. 

(iii) We assume that at fixed wavelength the effective noise can be
taken as a constant across the field of view, modulated only by the
number $n$ of independent exposures going into a given voxel (i.e.\
the exposure cube, see Figure \ref{individual-expmap}b):
$\sigma_\lambda^2 \approx s_\lambda^2 \times f_{\mathrm{rs},\lambda}
\times 4/n(x,y,\lambda)$. In other words, we assume the data to be
strictly background-limited and neglect the enhanced photon shot noise
in real objects for the estimation of the errors. While this implies
somewhat underestimated variances in the central pixels of bright
sources, it is an optimal assumption for faint objects where robust
error estimates are most important for detection and measurement
purposes.  

The 44 final released datacubes contained in the Data Release 1 (DR1)
have all been background-subtracted and the empirically
calculated effective noise has been inserted instead of the variance
noise scaled by the exposure cube. We release the individual MUSE-Wide
datacubes as a combined ``44-field'' datacube would have been too
large and inconvenient for further analysis both in terms of computer
memory and computation speed. 

\subsubsection{Estimation of the Point Spread Function in the final
  datacubes}\label{psf-estimation}

One important characterization of the datacubes is the estimation of
the point spread function (PSF). The Gaussian FWHM that the Autoguider
star measurement provides is only a rough approximation. Optimal
spectral extraction of compact sources requires good knowledge of the
MUSE PSF (see Section \ref{tdose}. Similarly, for the detection of
emission line sources using a matched filtering approach, we require
the PSF for cross-correlating with our model images (see Section
\ref{emdetect}). In addition, the large wavelength range covered by
MUSE required taking the variation of the PSF shape with wavelength
into account.   

The MUSE PSF has been shown\footnote{Tests on observations of Globular
  Clusters with MUSE show little discrepancy from the Moffat function
  accross the entire field of view \cite[e.g.,][]{husser16}.} to be
well characterized by a Moffat circular function \citep{moffat69}:

\begin{equation}
M(r) = \Sigma_0 \left( 1 + \left(\frac{r}{r_d}\right)^2
\right)^{\beta} \mathrm{,}
\end{equation}

\noindent
where $\Sigma_0$ denotes the central intensity, the width of the
profile is mainly determined by the dispersion radius $r_d$, while the
$\beta$-parameter defines the kurtosis of the profile. The full width
half maximum of the Moffat profile can then be expressed in terms of
$r_d$ and $\beta$ as $\mathrm{FWHM} = 2 \sqrt{2^{1/\beta}-1} r_d$

A theoretical description of wavelength dependence of the PSF
broadening has been derived in the framework of the Kolmogorov
turbulence model of the atmosphere \cite[e.g.,][]{tokovinin02}, but
for our purposes the decrease of the FWHM with wavelength can be
approximated with a linear function. We opted to keep the Moffat shape
parameter $\beta$ constant over the MUSE wavelength range; previous
experience has shown its variations to be negligible
\citep{pampelmuse}. We defined the reference wavelength to be at
7050~\AA, at the center of the MUSE wavelength range and for
comparison with the Autoguider measurement:  

\begin{equation}\label{fwhm-lambda}
\mathrm{FWHM}(\lambda)[\arcsec] = p_0 + p_1(\lambda - 7050\AA)
\end{equation}

In addition to the Moffat, we also computed the Gaussian profile of
the PSF, which misses its outer wings, but is in many ways easier to
handle. As for the Moffat, we assume circular symmetry for the
Gaussian representation, so that the PSF is fully described by its
FWHM, which also varies with wavelength according to Equation
\ref{fwhm-lambda}, albeit with different $p_0$ and $p_1$ factors.

We used altogether four different methods to estimate and model the
PSF in the combined datacubes, with some changes after completing the
first set of 24 fields. When multiple methods were available we always
selected the result that appeared most reliable, with the Gaussian
FWHM measurements obtained by the VLT Autoguider during the
observations as an additional independent check. 

\begin{itemize}

\item {\bf Method P}: Direct PSF fitting of stars in the field of view
  using PAMPELMUSE \citep{pampelmuse}. While a priori this seems the
  cleanest way to obtain the PSF, the stellar surface density in the
  CDFS is so low that less than 30\% of the fields contain at least
  one sufficiently bright star ($m_\mathrm{F814W} \la 22.5$). We
  modeled the PSF in MUSE collapsed mediumband images of 1150~\AA\
  width, i.e.\ for 4 wavelength bins, and then obtained the values of
  $p_0$ and $p_1$ of Eq.~\ref{fwhm-lambda} by fitting a linear
  function to the wavelength dependent FWHM.  

\item {\bf Method G}: Inferring the PSF from modeling compact
  galaxies. We visually selected from the HST/ACS F814W images
  relatively bright, compact galaxies without much structure, which we
  then convolved with a grid of different PSFs to match the MUSE
  resolution. The convolved and downsampled images were compared with
  MUSE collapsed mediumband images of 125~\AA\ width, and the
  best-match PSF parameters were then determined by minimizing
  $\chi^2$ over the grid, for each wavelength bin. $p_0$ and $p_1$
  were again obtained by fitting a linear function in wavelength. This
  method was only used to obtain Gaussian PSF parameters for fields
  01--24 and was later replaced by method C.  

\item {\bf Method C}: A hybrid method combining stars and compact
  galaxies. In order to go as faint as possible we modeled the PSF in
  only two broadband images corresponding to the HST/ACS bands F606W
  and F814W, which together cover the MUSE spectral range almost
  perfectly. Objects that proved difficult to model were excluded. The
  linear relation parameters $p_0$ and $p_1$ followed directly from
  the two broadband models. The method was applied to fields 25--46 to
  obtain both Gaussian and Moffat PSF parameters. 

\item {\bf Method F}: Full-frame modeling using the Fast Fourier
  Transform (FFT) method described in \cite{muse-deep}, applied to the
  comparison of {\it HST} and MUSE F606W and F814W broadband
  images. While potentially most powerful, this approach suffers from
  the need to exclude all stars with measurable proper motion between
  the {\it HST} and MUSE observation epochs (i.e.\ exactly those
  objects providing the best PSF constraints). We applied this method
  to estimate Moffat PSF parameters for all fields in this DR1, but in
  several cases (especially when there were stars in the field) the
  results from method C appeared more robust and were preferred.  
\end{itemize}

\begin{table*}
\caption{Moffat and Gaussian PSF parameters chosen to describe the PSF}\label{psftable}
\centering
\begin{tabular}{l c c c c c c c} 
\hline\hline
Field ID & $p_0$ Gaussian & $p_1$ Gaussian & Method\tablefootmark{a} & 
$p_0$ Moffat & $p_1$ Moffat & $\beta$ Moffat & Method\tablefootmark{a} \\
 & [\arcsec] & [10$^{-5}$\arcsec/\AA] & Gaussian & 
[\arcsec] & [10$^{-5}$\arcsec/\AA] & & Moffat \\
\hline\hline
candels-cdfs-01 & 0.836 & $-$4.429 & P & 0.744 & $-$3.528 & 2.800 & F \\
candels-cdfs-02 & 0.940 & $-$3.182 & G & 0.851 & $-$4.060 & 2.983 & P \\
candels-cdfs-03 & 0.944 & $-$4.460 & P & 0.809 & $-$4.060 & 2.859 & P \\
candels-cdfs-04 & 0.747 & $-$4.218 & G & 0.649 & $-$2.626 & 2.800 & F \\
candels-cdfs-05 & 1.025 & $-$3.003 & G & 1.148 & $-$4.693 & 2.800 & F \\
candels-cdfs-06 & 0.835 & $-$4.331 & G & 0.734 & $-$3.044 & 2.800 & F \\
candels-cdfs-07 & 0.935 & $-$3.966 & G & 0.871 & $-$3.663 & 2.800 & F \\
candels-cdfs-08 & 0.990 & $-$5.007 & G & 0.973 & $-$5.670 & 2.434 & P \\
candels-cdfs-09 & 0.832 & $-$8.069 & P & 0.726 & $-$2.180 & 2.857 & P \\
candels-cdfs-10 & 0.889 & $-$3.050 & G & 0.794 & $-$5.150 & 2.622 & P \\
candels-cdfs-11 & 0.988 & $-$3.770 & G & 0.934 & $-$5.410 & 2.815 & P \\
candels-cdfs-12 & 1.019 & $-$4.122 & G & 1.096 & $-$5.564 & 2.800 & F \\
candels-cdfs-13 & 1.063 & $-$5.284 & G & 1.166 & $-$5.332 & 2.800 & F \\
candels-cdfs-14 & 0.884 & $-$4.843 & G & 0.817 & $-$5.610 & 2.896 & P \\
candels-cdfs-15 & 0.702 & $-$4.441 & P & 0.735 & $-$4.840 & 2.917 & P \\
candels-cdfs-16 & 0.858 & $-$3.784 & P & 0.681 & $-$4.070 & 2.657 & P \\
candels-cdfs-17 & 0.780 & $-$3.534 & G & 0.644 & $-$4.690 & 2.245 & P \\
candels-cdfs-18 & 0.929 & $-$3.478 & G & 0.804 & $-$3.505 & 2.800 & F \\
candels-cdfs-19 & 0.814 & $-$3.524 & G & 0.676 & $-$2.603 & 2.800 & F \\
candels-cdfs-20 & 0.712 & $-$5.196 & G & 0.670 & $-$5.006 & 2.800 & F \\
candels-cdfs-21 & 0.835 & $-$4.255 & P & 0.598 & $-$2.930 & 3.029 & P \\
candels-cdfs-22 & 0.787 & $-$3.252 & P & 0.725 & $-$5.710 & 3.078 & P \\
candels-cdfs-23 & 0.777 & $-$3.018 & G & 0.720 & $-$3.818 & 2.800 & F \\
candels-cdfs-24 & 0.728 & $-$4.232 & G & 0.634 & $-$3.190 & 2.800 & F \\
candels-cdfs-25 & 0.830 & $-$4.310 & C & 0.696 & $-$3.590 & 3.108 & P \\
candels-cdfs-26 & 2.002 & $-$2.630 & C & 1.672 & $-$6.483 & 2.800 & F \\
candels-cdfs-28 & 0.963 & $-$4.439 & --- \tablefootmark{b} & 0.830 & $-$2.667 & 2.800 & F \\
candels-cdfs-29 & 1.060 & $-$3.550 & C & 0.955 & $-$3.190 & 3.266 & P \\
candels-cdfs-30 & 1.165 & $-$5.690 & C & 1.021 & $-$4.370 & 3.681 & P \\
candels-cdfs-31 & 0.969 & $-$4.450 & C & 0.794 & $-$4.600 & 2.593 & P \\
candels-cdfs-32 & 1.275 & $-$5.980 & C & 0.968 & $-$0.224 & 2.800 & F \\
candels-cdfs-33 & 1.292 & $-$4.900 & C & 1.137 & $-$5.140 & 2.800 & F \\
candels-cdfs-34 & 1.286 & $-$8.770 & C & 0.941 & $-$4.821 & 2.800 & F \\
candels-cdfs-35 & 0.736 & $-$2.840 & C & 0.656 & $-$3.052 & 2.800 & F \\
candels-cdfs-36 & 0.833 & $-$3.860 & C & 0.758 & $-$4.100 & 2.293 & P \\
candels-cdfs-37 & 0.884 & $-$2.920 & C & 0.717 & $-$4.109 & 2.800 & F \\
candels-cdfs-39 & 0.753 & $-$3.420 & C & 0.638 & $-$3.430 & 2.333 & P \\
candels-cdfs-40 & 1.336 & $-$6.290 & C & 1.124 & $-$4.575 & 2.800 & F \\
candels-cdfs-41 & 1.097 & $-$5.160 & C & 0.880 & $-$2.270 & 2.696 & P \\
candels-cdfs-42 & 1.380 & $-$5.840 & C & 1.190 & $-$6.794 & 2.800 & F \\
candels-cdfs-43 & 1.217 & $-$3.640 & C & 1.010 & $-$1.526 & 2.800 & F \\
candels-cdfs-44 & 0.971 & $-$6.280 & C & 0.849 & $-$3.528 & 2.800 & F \\
candels-cdfs-45 & 0.893 & $-$3.140 & C & 0.756 & $-$3.140 & 3.119 & P \\
candels-cdfs-46 & 0.859 & $-$4.610 & C & 0.707 & $-$2.990 & 2.279 & P \\
\hline
\end{tabular}
\tablefoot{
\tablefoottext{a}{Methods (as described in text): P=direct fitting of
  stars, G=modeling of compact galaxies, C=hybrid method of fitting
  stars and galaxies combined, F=Fast Fourier Transfrom}
\tablefoottext{b}{Fitting resulted in positive slope $p_1$ - use a
  mean of slopes of other fields with similar Airmass and Autoguider
  seeing and fix FWHM to Autoguider value.}}
\end{table*}

\begin{figure}
\centering
\includegraphics[width=8.8cm]{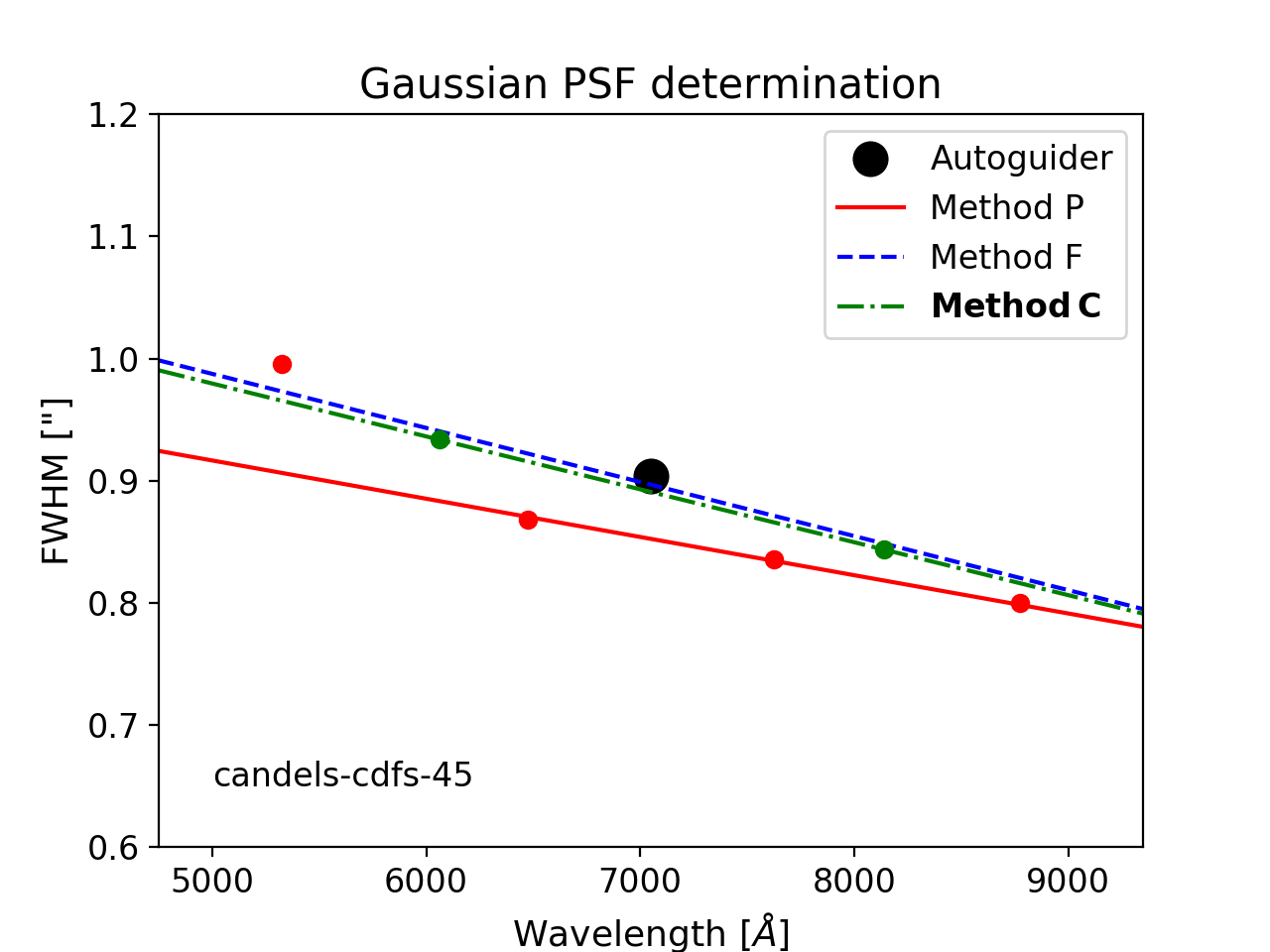}
\includegraphics[width=8.8cm]{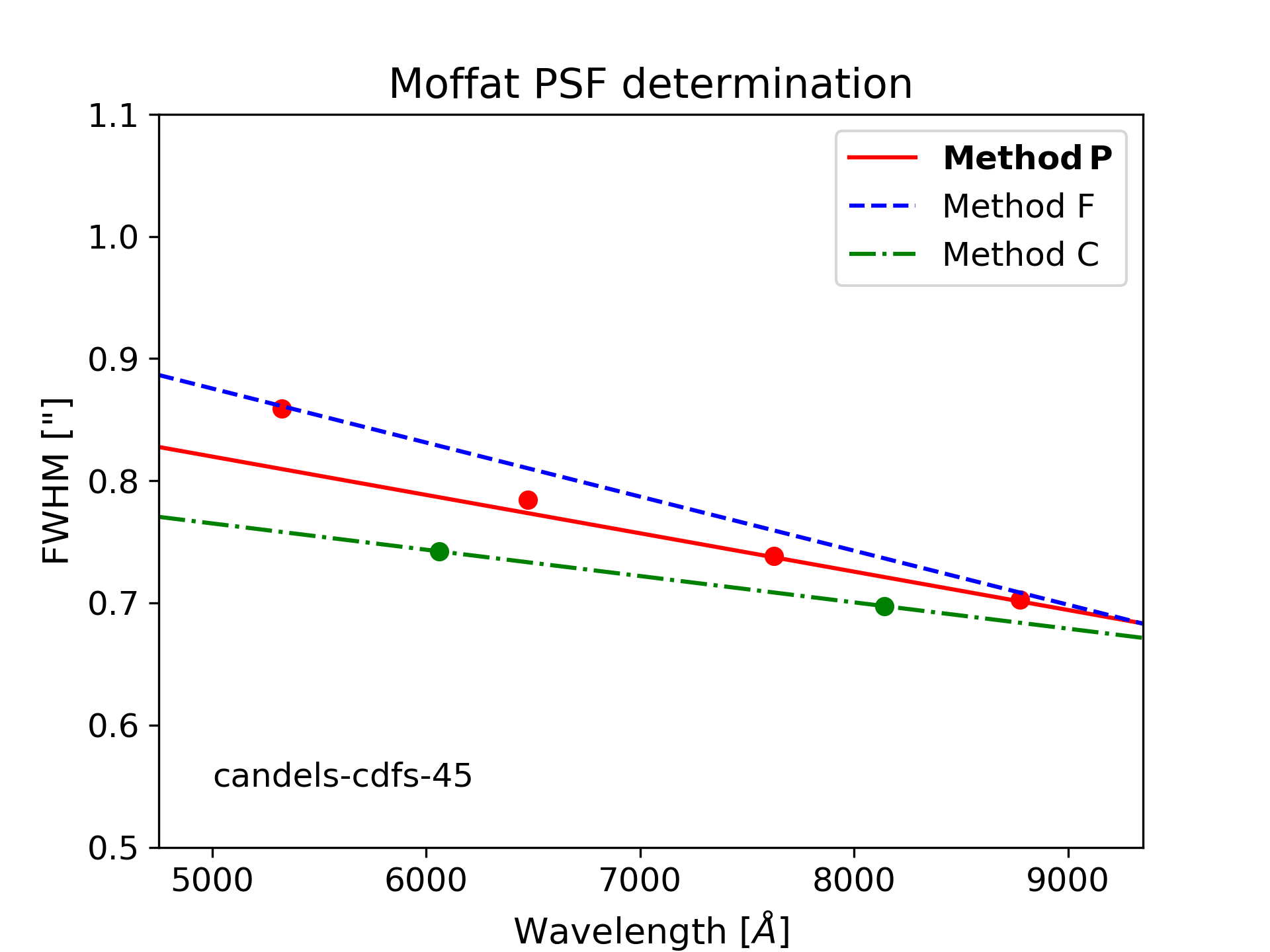}
\caption{Example of determined PSF slopes for the various methods on
  the field candels-cdfs-45: in the upper panel for the Gaussian $p_0$
  and $p_1$ values and in the lower for the Moffat values. Methods (as
  described in text): P=direct fitting of stars, C=hybrid method of
  fitting stars and galaxies combined, F=Fast Fourier Transfrom. The
  final selected method is marked in bold letters.}\label{psf-slopes} 
\end{figure}

Table \ref{psftable} documents which method was finally used for which
parameter set. Figure \ref{psf-slopes} shows an example of the
different FWHM determined as a function of wavelength for both the
circular Moffat and Gaussian parameters. Similar figures of the PSF
determination are found in the Quality Control pages of the data
release webpage for each field (see Appendix \ref{webpage}). 

\section{Emission line selected objects}\label{emline}

MUSE can efficiently locate sources in a 3D cube without any
photometric pre-selection; it is particularly powerful for the
detection of emission line objects. Most of the science cases of
MUSE-Wide rely on emission line sources detected in a homogeneous
manner (Section \ref{sciencecase}). In \citetalias{herenz17} we
provided a catalog of 831 emission line sources for the first 24
fields of MUSE-Wide using a matched filtering approach. Since then we
employed the same strategy for the other 20 fields of the MUSE-Wide
DR1. Here we only provide a brief description of the process already
described in \citetalias{herenz17}.  

\subsection{Detection and classification}\label{emdetect} 

Prior to searching for emission lines in the datacube we had to remove
any underlying continuum signal from the specta. To achieve this goal
we subtracted a 151 pixels wide running median in spectral direction
from the datacube. The continuum-subtracted cube was then fed into the
LSDCat software \citep{lsdcat}, together with the empirically
determined ``effective variances'' (Section \ref{effectivenoise}). In
brief, LSDCat cross-correlates the entire cube with a 3D source
template and provides a list of emission line detections graded by
significance.  For the spatial template we adopted for each field a
circular Gaussian with a FWHM of the PSF (Section
\ref{psf-estimation}), thus targeting in particular compact emission
line sources. For the spectral template we took again a Gaussian, but
with a FWHM fixed in velocity space to a value of 250~km/s, a value
optimised to find \Lya\, emitters. However, the algorithm is quite
robust against template mismatches\cite[see discussion in][]{lsdcat}. 

\begin{figure*}[ht!]
\centering
\includegraphics[width=9.1cm]{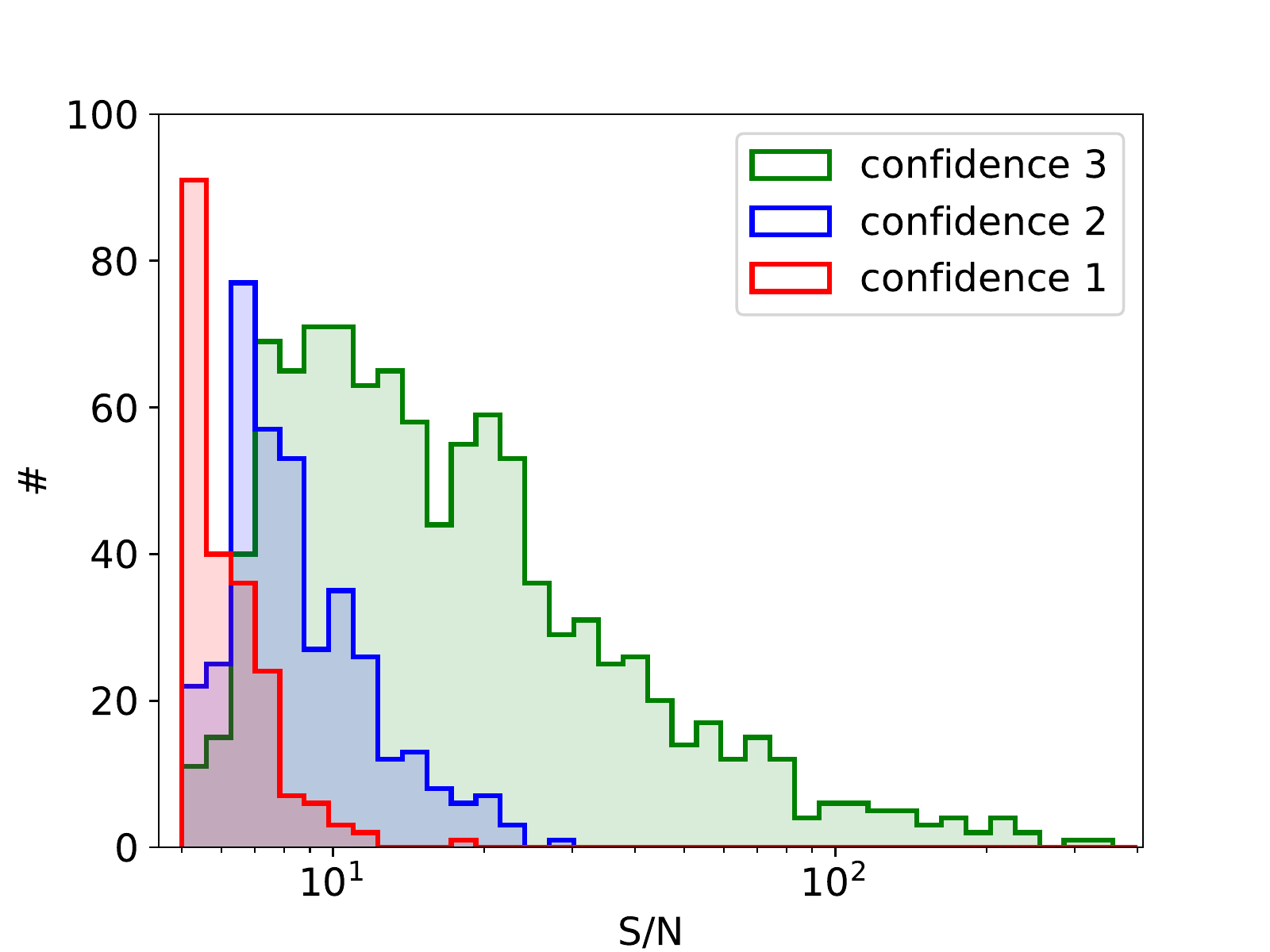}
\includegraphics[width=9.1cm]{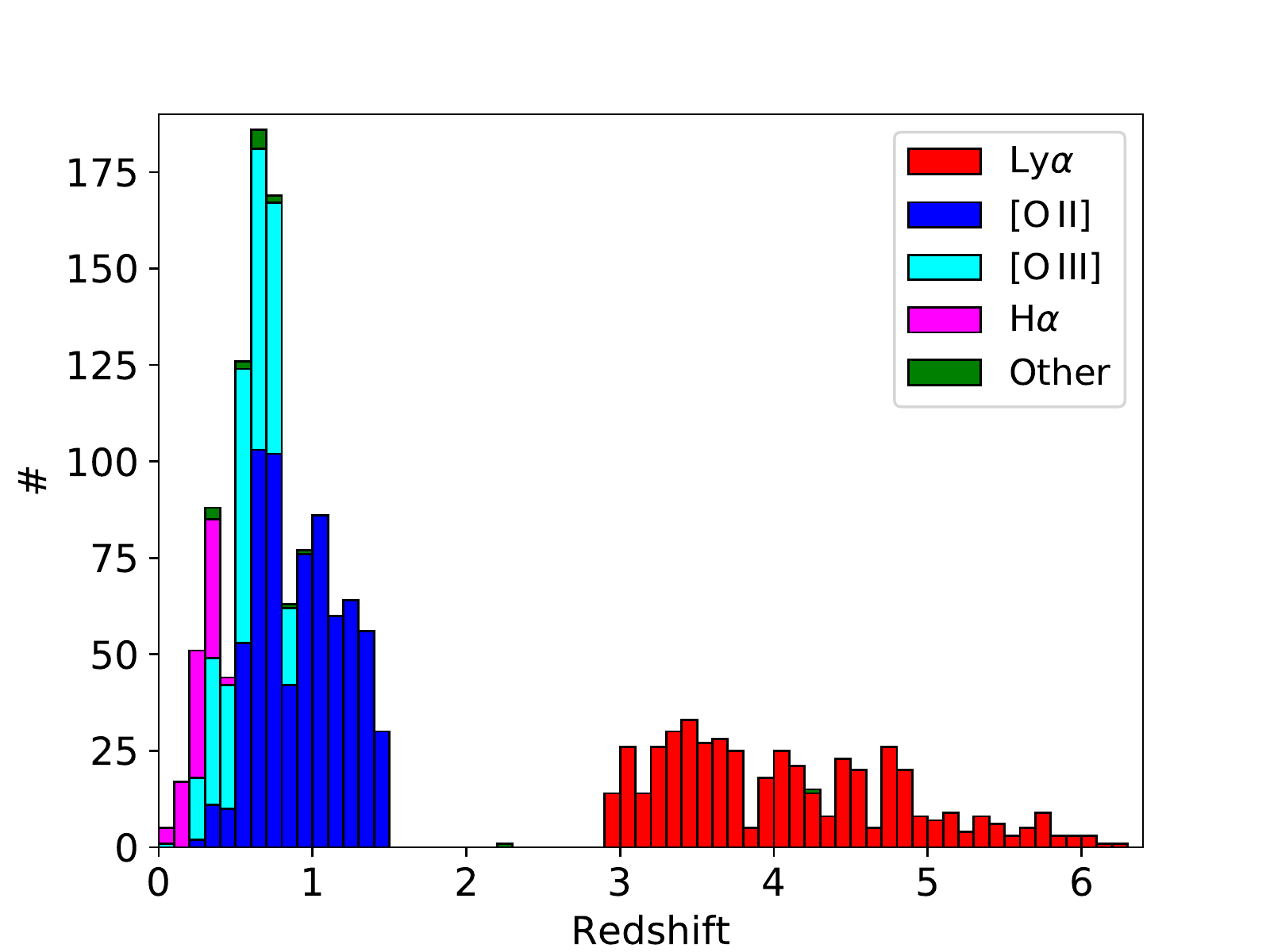}
\caption{(a) Left: Distribution of number of emission-line objects as
  a function of S/N of the lead line for each of the three confidence
  levels we employ for classification. Objects at the S/N limit are
  dominated by confidence 1 objects and may include some spurious
  lines, while for objects with S/N above 10, we expect hardly any
  misclassifications. (b) Right: Redshift histogram of the emission
  line sources classified by their strongest line. The redshift
  desert, where there are no strong emission lines in the MUSE
  wavelength range between 1.5 $<$z$<$2.9, is clearly visible. We are
  able to find 479 LAEs, reaching densities of almost 12
  LAEs/arcmin$^2$.} 
\label{em-redshift-sn}
\end{figure*} 

In order to define candidate emission line detections, LSDCat requires
a detection threshold in the signal-to-noise ratio (S/N). In
comparison to \citetalias{herenz17} we lowered this S/N threshold to a
value of 5, which turned out to be as low as we could go before
getting strongly affected by spurious detections (see below). We
note that this lower threshold applies only to the newly added fields,
while the detection limit for the first 24 fields was 8 in S/N using
the old recipe for the effective variances, which converts into a
value of 6.4 with the new improved prescription\footnote{This
  difference in the effective noise scale can be explained by the fact
  that the ``old'' recipe as used in \citetalias{herenz17} did not
  correct for the spectral resampling; this amounts to a factor 1.25
  in the noise level, for standard MUSE settings.}. On output, LSDCat
groups multiple line detections together that were found within a
certain radius (which we set to $0.8\arcsec$). A candidate object thus
consists of one or several detected lines, where the detection with
the highest S/N is denoted as ``lead line''.  

In the next step we visually inspected and classified all detected
objects with our QtClassify
tool\footnote{\url{http://ascl.net/1703.011}, see Appendix A of
  \citetalias{herenz17} for a description of the functionality}
\citep{qtclassify} in a two stage process: In a first pass, each
object was classified independently by two team members, followed by a
second pass where these two and a third member as referee had to agree
on the final classifications. The referee had final say on cases in
which the initial classifiers disagreed. During the inspection process
we purged spurious sources such as sky or continuum subtraction
residuals, classified the remaining objects by identifying the lead
line (and thus setting the redshift), and assigned qualitative
indicators describing the robustness of the classification. For the
latter we distinguish between ``quality'' and ``confidence'': Quality
specifies whether any secondary lines have aided the classification
process (``A'' for multiple lines above the S/N threshold, ``B'' if
only one line was detected, but more are visible in an extracted S/N
spectrum and match in redshift, and ``C'' for single-line
objects). The ``confidence'' value is a more subjective interpretation
of our trust in the classification, with a value of 3 expressing very
high certainty, 2 representing a still quite trustful result (expected
error rate $\lesssim 10\%$), and 1 the lowest confidence with an
assumed error probability in the correct identification of the line of
up to $\approx 50\%$. Also here the referee in the second
classification pass had final say on the quality and confidence
indicators in the catalog, especially when the two initial classifiers
did not agree. Even with multiple classification passes, some degree
of subjectivity remained, particularly at the boundaries.

We emphasize that the leading emission line detection of confidence 1
objects is still highly significant, and we expect a low rate of
entirely spurious detections. Comparison of the MUSE-Wide emission
line catalog in the UDF with the MUSE-Deep catalog utilizing the full
10 hour exposure time \citep{inami17}, yielded $\approx 5\%$ false
positives and all were at a S/N level less than 6. We will explore the
comparison of the shallow MUSE-Wide survey versus MUSE-Deep further,
when we release that portion of MUSE-Wide in a future Data Release
2. The lower confidence mainly reflects the ambiguity of the line
identification, not the fidelity of the source itself. Figure
\ref{em-redshift-sn}a shows the distribution of S/N values of the lead
emission line for the three different confidence levels; our
confidence level clearly depends strongly on the S/N of the lead line.  

While LSDCat and QtClassify already yielded provisional redshifts,
these were subsequently refined as follows: We extracted PSF-weighted
one-dimensional (1D) spectra at the position of the emission line
source, with both air and vacuum wavelengths using the Vienna atomic
line database formalism \citep{air2vac}%
\footnote{http://www.astro.uu.se/valdwiki/Air-to-vacuum\%20conversion}.
\Lya-based redshifts were then based on the peaks of fitted line
profiles assuming an asymmetric Gaussian line shape \citep{shibuya14},
with however no correction for any offset the \Lya\, line from
systemic. Redshifts for other emission line galaxies ($z < 1.5$) were
determined by fitting Gaussian line profiles simultaneously to all
emission lines present in the object. For the \OII\, doublet we used a
double Gaussian with fixed separation, all other lines were fitted
with single components. The final redshift was taken as the
S/N-weighted mean of all lines, and redshift errors were estimated by
repeating the fitting procedure 100$\times$ on the spectra after
randomly perturbing them according to the effective noise.  

\begin{figure*}
\centering
\includegraphics[width = 14cm]{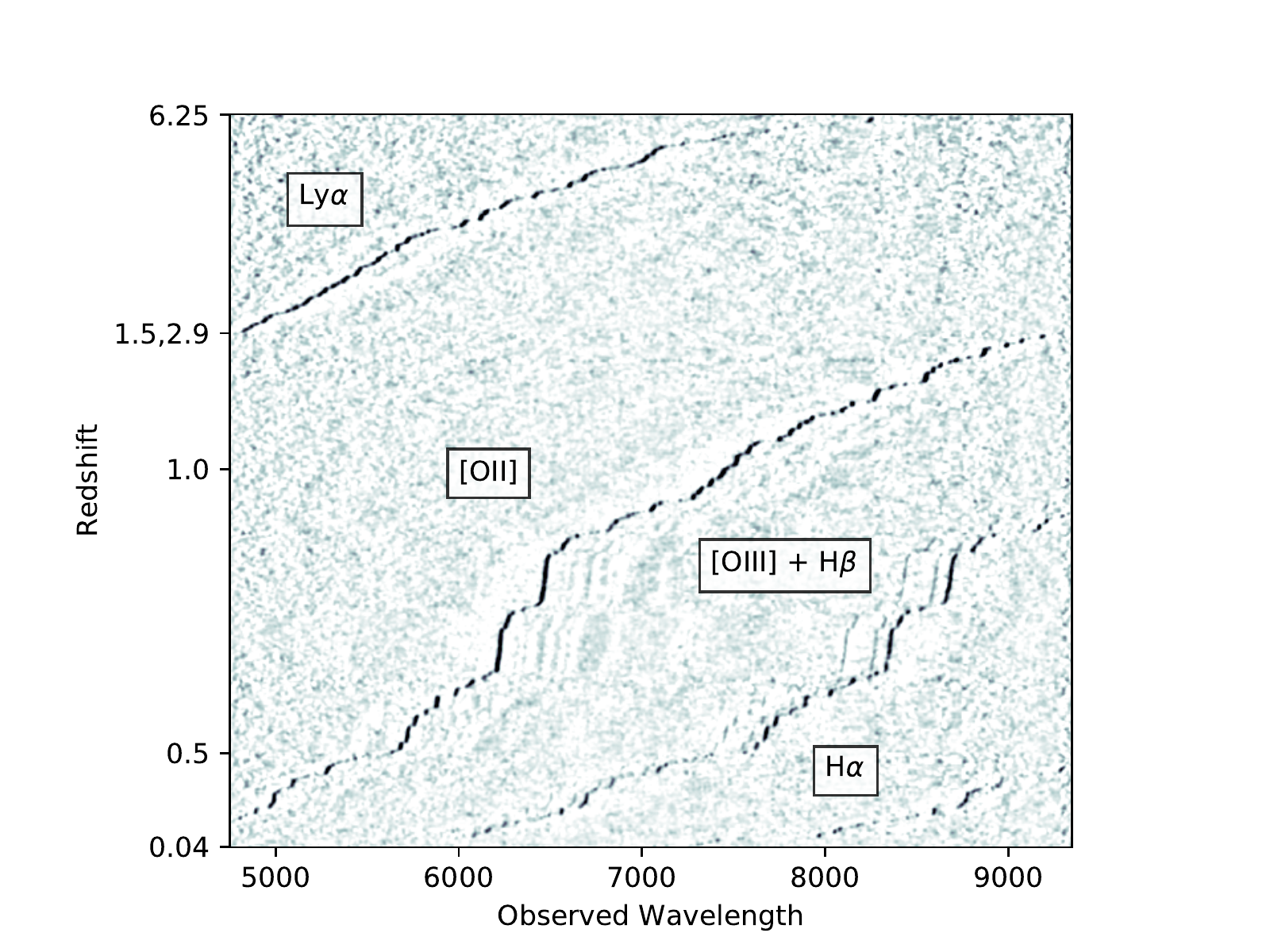}
\caption{Stack of normalized spectra of the emission line
  objects. They are stacked in $y$-direction with increasing redshifts,
  with a large jump between z$\sim$1.5 and z$\sim$2.9. First we
  normalized the spectra to the brightest emission line, then we
  smoothed with a 10~\AA\, Gaussian and finally we smoothed the 2D image
  with a 2.8 pixel 2D circular Gaussian.}\label{emlinestack} 
\end{figure*}

Following classification we created a merged object catalog for the
entire DR1 footprint. We discarded double detections in overlapping
regions of adjacent MUSE-Wide fields (always retaining the detection
with higher S/N). We also had to perform some manual interventions
such as grouping emission line sources belonging to the same galaxy,
but which was too extended for the automatic grouping of LSDCat, and
splitting up superpositions of different-redshift emission line
objects closely aligned in the line of sight. After merging and
cleaning we were left with a final catalog of 1,602 emission line
objects, based in 3,057 detected emission lines. The redshift
distribution of the objects, grouped by their lead-line
identifications, is shown in Figure \ref{em-redshift-sn}b. This plot
shows a clear redshift desert between $z\simeq 1.5$ (where \OII\, is
redshifted out of MUSE) and $z\simeq 2.9$ (where \Lya\, enters); the
region in between is populated only by two AGN. We note that the
continuum-selected sample discussed in Section \ref{photocat} does not
have such a redshift desert (see also Figure
\ref{redshift-photo}). Figure \ref{emlinestack} shows a montage of all
1602 emission line object spectra stacked in $y$ direction with
increasing redshifts.   

The released data tables (object catalog and emission line table) are
described in Section \ref{emcats} below. Here we briefly introduce the
unique identifiers of MUSE-Wide emission line objects, UNIQUE\_ID in
the catalogs. It is composed of 9 digits and divided into 4 groups in
the format ``ABBCCCDDD''. The first digit refers to one of the five
parent regions in which the MUSE pointing was obtained (which is
always 1 in DR1, for candels-cdfs). Next comes the two-digit number
characterising the field in which the object was discovered. CCC
refers to the LSDCat object identifier in that field, and DDD refers
to the emission line running number for the lead line in the emission
line table. The last number is important for distinguishing
superpositions of objects at different redshift that were assigned the
same object ID by LSDCat. Thus for example, the source with unique
identifier 106043096 was found in field candels-cdfs-06 as LSDCat
object 43, and its lead line has the running number 96 in the emission
line table.

\subsection{Cross-match with photometric and spectroscopic catalogs}\label{crossmatch}

By cross matching the MUSE emission line objects to multiband {\it
  HST} catalogs we could add broad band photometry, especially far
into the Near-IR wavelengths unaccessible with MUSE. For the
cross-match we use two catalogues currently available in the
CANDELS/CDFS region: the \cite{guo} CANDELS catalog based on deep
F160W WFC3 imaging and the \cite{skelton} 3D-HST catalog based on a
combination of the F125W/F140W/F160W WFC3 filters. While the 3D-HST
catalog is deeper, it shows higher fragmentation of sources at low
redshift. Also, while the 3D-HST offers excellent and vast photometry,
particularly in the Near-IR data, the CANDELS catalog provides more
complete links to other multiwavelength information, such as X-ray and
radio. 

We determined the photometric counterparts to our emission line
sources by searching for the nearest counterpart within
$0.5\arcsec$. In \citetalias{herenz17} we had estimated the 3$\sigma$
positional error between the {\it HST} catalogs and the MUSE LSDCat
position of the emission lines to be $<0.5\arcsec$. We then visually
inspected the {\it HST} image cutouts and consolidated the counterpart
list, either by adding potential counterparts outside $0.5\arcsec$ or
by purging the closest catalogued counterpart if it did not match our
expectations of the emission line (e.g. no drop in the broad band
images representing the rest-frame Lyman continuum for a
\Lya-emitter).     

Table \ref{counterparts} shows the percentages of MUSE counterparts
found in the two photometric catalogs. As expected, we are nearly
complete at low redshift, the main source of incompleteness being
superpositions with large galaxies. At high redshift, the LAEs show a
much lower percentage of photometric counterparts. In some of those
sources we find a clear counterpart in the optical {\it HST} images,
but they are not catalogued in the near-IR selected catalogs, possibly
because of their UV-dominated spectra energy distribution (SED). Other
high redshift sources are just below the detection limit of the
broad-band images, hence the higher percentages of LAE counterparts in
the deeper 3D-HST catalog. As in the MUSE deep fields, also here we
detect several LAEs without any photometric counterparts, neither in
the images nor catalogs \citep[Maseda et al. submitted]{muse-deep};
these constitute some of the highest equivalent width sources known
(EW$_0 >$ 500\AA) and will be the subject of further study within
MUSE-Wide (Kerutt et al. in prep.). Lastly, some of the LAEs without
counterparts at very low S/N and low confidence may be spurious
detections within MUSE and not real sources, but we estimate that
fraction to be $\lesssim 5\%$.

\begin{figure*}
\centering
  \includegraphics[width=14.0cm]{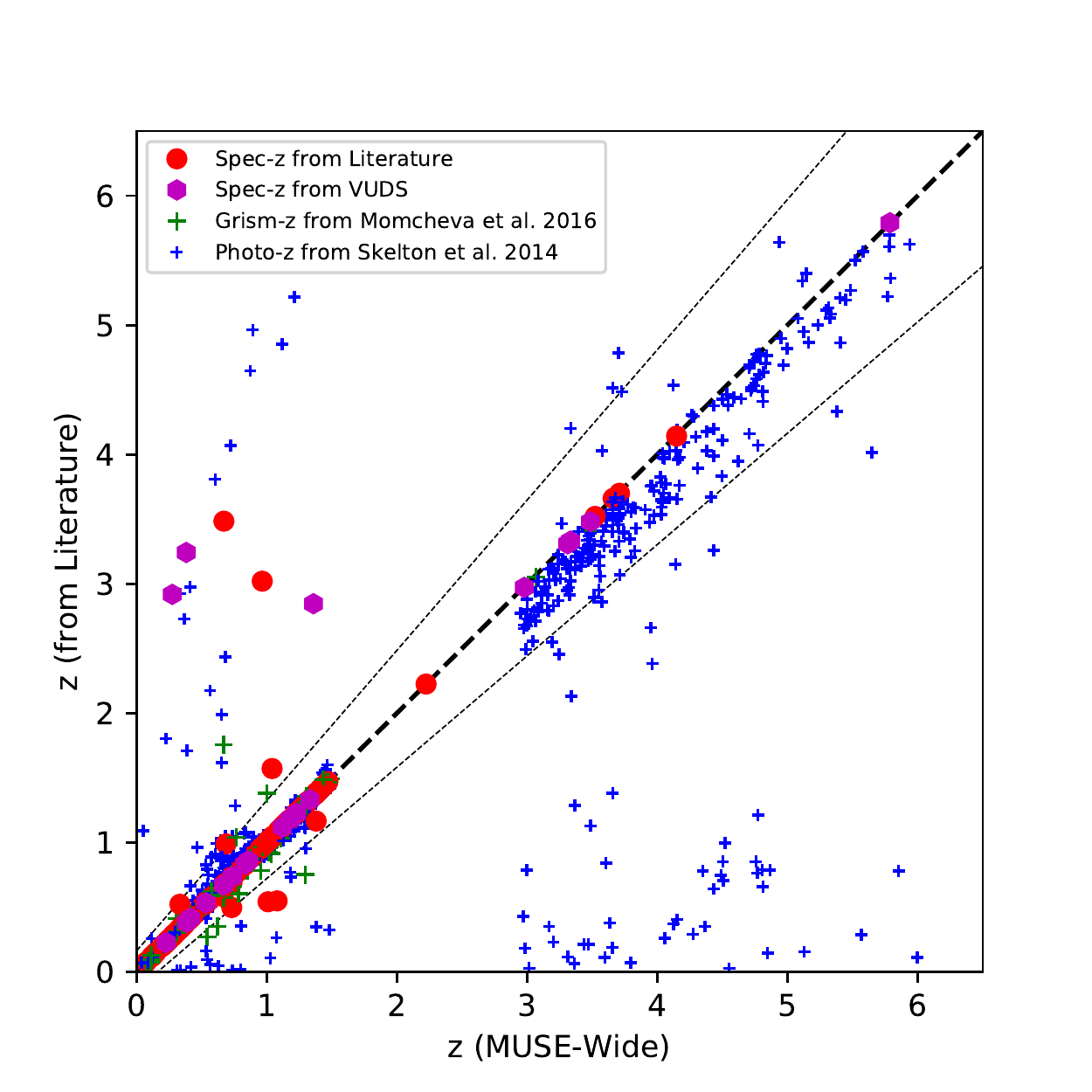}
    \caption{Redshift comparison between MUSE and literature redshifts
      for our emission line selected galaxies. Red dots denote the
      compilation of redshifts from the literature gathered in
      \cite{skelton}. Magenta hexagons represent spectroscopic
      measurements from VUDS \citep{tasca17}. Green crosses represent
      grism spectroscopic redshifts from \cite{momcheva16}, which were
      aided by the photometric redshifts from \cite{skelton} denoted
      with blue crosses. The thin dashed lines show the regions
      outside of which a photometric redshift is determined as a
      catastrophic failure.}\label{photz-compare} 
\end{figure*}

\begin{table}[ht!]
\caption{Counterpart percentages between the emission line sources to
  {\it HST} selected catalogs}\label{counterparts}
\centering
\begin{tabular}{c c c c} 
\hline\hline
Photometric & & LAEs & Total \\
catalog & $z < 2.9$ & $z > 2.9$ & galaxies \\
\hline\hline
\cite{guo} & 1064 (95\%) & 212 (44\%) & 1276 (80\%) \\
\cite{skelton} & 1083 (96\%) & 272 (57\%) & 1355 (85\%) \\
\hline
\end{tabular}
\end{table}

We also compared the 1,355 emission line galaxies with 3D-HST
\citep{3dhst} counterparts to their redshift measurements from the
literature. Although the CANDELS team has an internal photometric
redshift selection combining 6 photometric redshift (photo-$z$) codes
following the study of \cite{dahlen}, only one of those photometric
redshift determinations is public \citep{hsu14}. We opted to wait
until the CANDELS collaboration releases their photometric redshifts to
compare the MUSE spectroscopic redshifts to them. The bulk of the
3D-HST redshift measurements came from \cite{skelton} using the EAzY
code \citep{eazy} to determine their photometric redshifts. EAzY
benefits from the large amount of photometric bands that 3D-HST
provides. Furthermore \cite{skelton} included previous spectroscopic
redshifts from their study (see \cite{wuyts08} for a compilation of
the different spectroscopic campaigns used). We added 330 mostly low
redshift sources with updated {\it HST} grism spectroscopy from
\cite{momcheva16}. The changes between the \cite{skelton} photometric
redshifts and the grism redshifts are minimal, since the grism
identification is aided (and often dominated) by the photo-$z$. We
furthermore added 22 sources with new spectroscopy from the VIMOS
Ultra-Deep Survey (VUDS) \citep{lefevre15,tasca17} to our comparison
for a total of 307 objects with spectroscopic redshifts (not counting
the 330 grism redshifts).  

Figure \ref{photz-compare} shows the comparison between our
spectroscopic classification and various redshift values from the
literature, including a majority of photometric redshifts from
\cite{skelton}. There exists a systematic offset between the
photometric and spectroscopic redshifts for our high redshift LAEs
when there is not a catastrophic redshift failure, with the median
offset between the MUSE redshift and the 3D-HST photo-$z$ being
$\Delta$z $\sim$ 0.2. This offset has been remarked upon by
\cite{oyarzun16} and most likely relates to EAzY shifting a slightly
blueshifted Lyman break when strong \Lya\, emission is present to
account for the flux excess in the redder band. An extensive
investigation into the sources of mismatch between MUSE spectroscopic
redshifts and photometric redshifts was studied by the MUSE-Deep
survey \citep{brinchmann17}. In addition to the template mismatch in
EAzY noted above, the authors explain that a source of further
contribution to the offset at high redshift relates to the amount of
intergalactic absorption these high redshift galaxies experience. They
also find, perhaps counterintuitively, that adding extensive ground
and mid-IR photometry to very faint sources worsens the photo-$z$
prediction. Lastly, they remark that a wrong association can play a
significant role, which we also find when comparing our sources with
spectroscopic samples below.

\begin{table*}[ht!]
\caption{Catastrophic failures for objects with spectroscopic
  redshifts from the literature}\label{specz-fail}
\centering
\begin{tabular}{c c c c c c l } 
\hline\hline
ID  & ID & $z$ & $z$ & quality & $z_{\mathrm{MW}}$ & comment \\
MUSE-Wide & 3D-HST & MUSE & Literature & confidence & correct? & \\
\hline
106043096 & 23930 & 1.080 & 0.549 & b-3 & yes & 
clear \OII\, doublet, $z_{\mathrm{Lit.}}$ misclassified as \OIII \\
108022137 & 16741 & 0.732 & 0.497 & b-3 & yes & 
aided by 4000~\AA\, jump \\
112008041 & 20496 & 0.671 & 3.484 & c-2 & yes & 
superposition, $z_{\mathrm{Lit.}}$ refers to CANDELS \# 20768 \\
115018112 & 09759 & 0.966 & 3.020 & c-3 & yes & 
affected by extended \Lya\, from 115004089 \\
119035074 & 16008 & 1.041 & 1.572 & c-2 & yes & 
doublet and 4000\AA\, jump \\
123045174 & 16325 & 0.688 & 0.989 & a-3 & yes & 
4 extra emission lines to aid classification \\
137012028 & 33145 & 0.382 & 3.242 & c-2 & no & 
probably misclassified line as \OII\, by MUSE-Wide \\
140002014 & 32236 & 0.334 & 0.523 & a-3 & yes & 
5 extra emission lines to aid classification \\
141056169 & 27980 & 1.380 & 1.166 & c-1 & yes & 
only conf. 1, but photo-$z$ agrees with MUSE-Wide \\
142049165 & 30364 & 0.275 & 2.918 & b-3 & yes & 
literature spectrum probably misclassified \OII\, for \Lya\\
145049108 & 25822 & 1.010 & 0.542 & c-1 & yes & 
only conf. 1, but photo-$z$ agrees with MUSE-Wide \\
146080366 & 29021 & 1.359 & 2.846 & c-2 & yes & 
photo-$z$ agrees with MUSE-Wide \\
\hline
\end{tabular}
\end{table*}

We defined a catastrophic redshift failure between literature and MUSE
redshifts to occur if the following condition was met for photometric
redshifts: 

\begin{equation}
\left| \ln \frac{(1+z_{\mathrm{phot}})}{(1+z_{\mathrm{MUSE}})} \right| > 0.15
\end{equation}

\noindent
and $|z_{\mathrm{spec}} - z_{\mathrm{MUSE}}| > 0.1$ for spectroscopic
redshifts. 113 objects (8\% of 1,355 emission line objects) satisfy
those conditions, with the majority (101) coming from catastrophic
photometric redshift errors. We investigate the 12 mismatches between
literature spectroscopic redshifts and MUSE spectroscopic redshifts in
Table \ref{specz-fail}. Except for object 137012028, which shows a
photo-$z$ probability distribution function matching better to the
high redshift solution, we are quite certain in our classifications of
the sources, often having other lines or spectral features to aid us
in our assessment of the redshift.

\subsection{Stellar Masses}\label{masses}

We also determined the stellar masses of the emission line objects
with a catalogued photometric counterpart. SED
derived stellar masses carry many uncertainties, e.g. the proper
accounting for emission lines. We caution that the stellar masses
derived serve only as an estimate. We used the \cite{skelton}
photometry and the software FAST \cite[Fitting and Assessment of
Synthetic Templates,][]{fast} similar to the 3D-HST team, allowing for
an easy comparison between the samples. FAST determines the best fit
parameters using $\chi^2$ minimization from a set of model SEDs and an
analysis grid describing several stellar population models. 

The stellar population model grid parameters are: stellar age,
characteristic star formation timescale $\tau$, dust content A$_V$,
metallicity and redshift (which we fixed to the MUSE redshift). As
in \cite{skelton}, we employed the \cite{bc03} stellar library, the
\cite{chabrier} initial mass function and the \cite{calzetti} dust law
for our fits. For low-redshift sources ($z<2.9$) we used an
exponentially declining star formation history, where $\tau$ refers to
the width of the declining exponential, while the stellar age is when
the star-formation burst happened before the exponential
decline. However, for our high redshift LAEs this model is likely
unphysical, since the galaxies are going through a young burst, which
is possibly their first and dominates the continuum of the
sources. Using a truncated star formation history in which $\tau$
corresponds to the length of the burst and is equivalent to the age
improved the $\chi^2$ of the fits dramatically, even if they could not
capture ages below 40 Myr in the models. We expanded and refined the
analysis grid of \cite{skelton} slightly, for example employing finer
steps for the stellar ages, dust attenuations and using 4
metallicities (0.004, 0.008, 0.02 and 0.05 Z$_{\odot}$) instead of
just one (0.02 Z$_{\odot}$) used by \cite{skelton}. Once a preferred
stellar model was found, stellar masses were derived via the mass to
light ratios of that model adjusted to the SED photometry. 

\begin{figure}
\centering
\includegraphics[width=9.2cm]{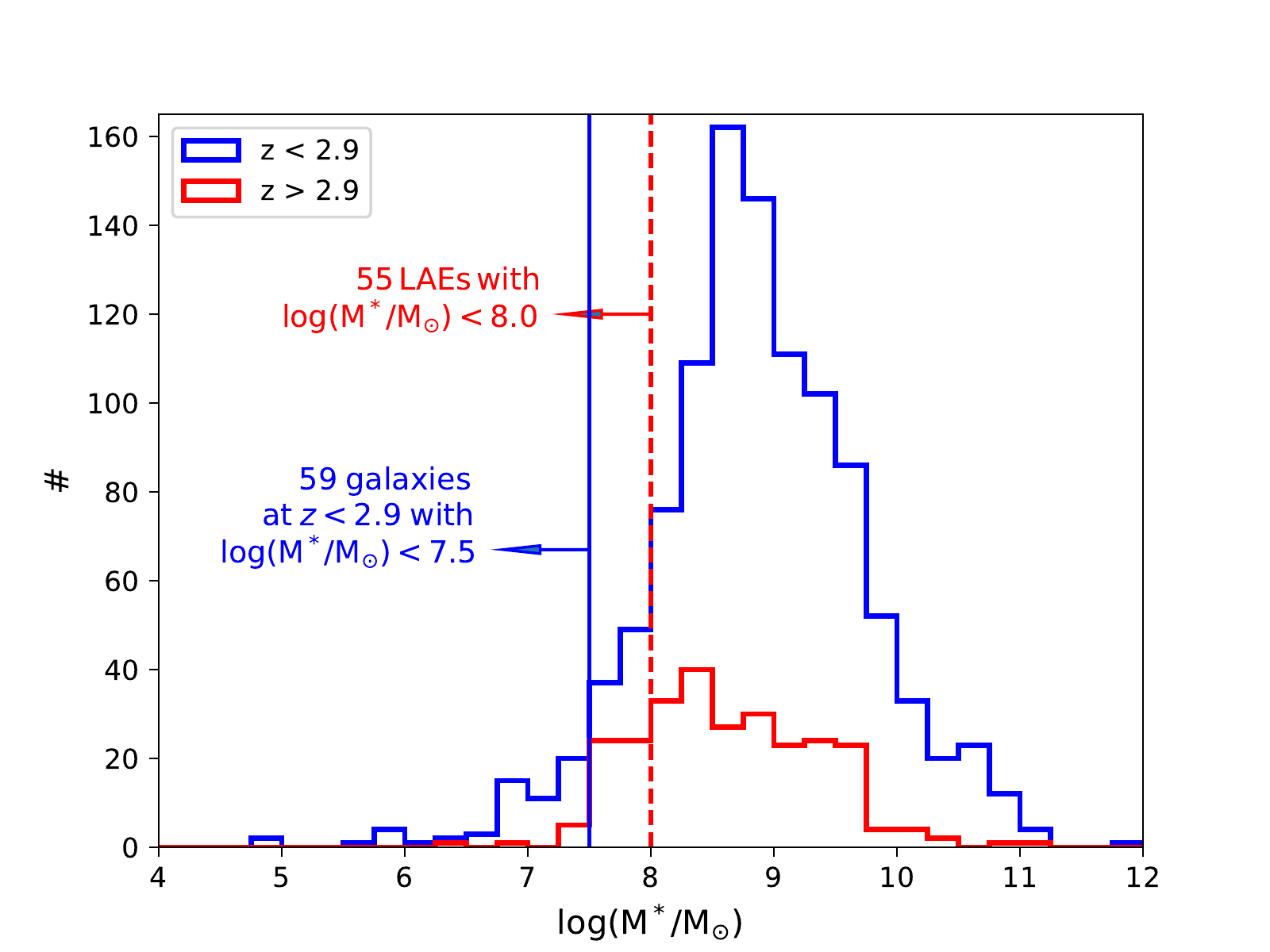}
\caption{Stellar mass histogram for emission-line objects with
  photometric counterparts, in red for high-redshift LAEs, in blue for
lower redshift objects. We find a tail of low mass dwarf galaxies
representing some of the lowest mass galaxies known at these
redshifts.}\label{stellarmass}
\end{figure}

Figure \ref{stellarmass} shows the distribution of the thus obtained
stellar masses of emission line galaxies, divided into high and low
redshift subsets. Both distributions are fairly broad, but show a
significant tail towards low mass, dwarf systems; 55 LAEs have stellar
masses lower than log(M$^*$/M$_{\odot}$) $=8.0$, while 59 intermediate
redshift galaxies have masses lower than
log(M$^*$/M$_{\odot}$) $=7.5$. The LAEs typically have lower masses
than other galaxies, such as Lyman Break Galaxies (LBGs) found at
these redshifts. However, this is just a consequence of the selection
method; the need for photometry skews them to have higher stellar
masses \citep{hagen16}. At intermediate redshifts, MUSE-Wide is able
to peer into the field population of star forming dwarfs. If emission
lines were included in the SED fitting the steller masses would likely
decrease for both samples, further emphasizing that this emission line
catalog skews towards low mass systems.

We do not provide age estimates, dust attenuation or star formation
rates (SFR) derived from the stellar models as these SED derived
parameters have been shown to contain large uncertainties and should
be dealt with on a case-by-case basis taking into account emission
line strengths for their analysis \cite[e.g.,][]{stark13}. On average,
though, LAEs are known to have extremely young populations with a
large fraction hitting the 40 Myr limit and specific star formation
rates lying well above the high redshift main sequence
\citep{speagle14}. 

\begin{figure*}[ht!]
\centering
\includegraphics[width=9.0cm]{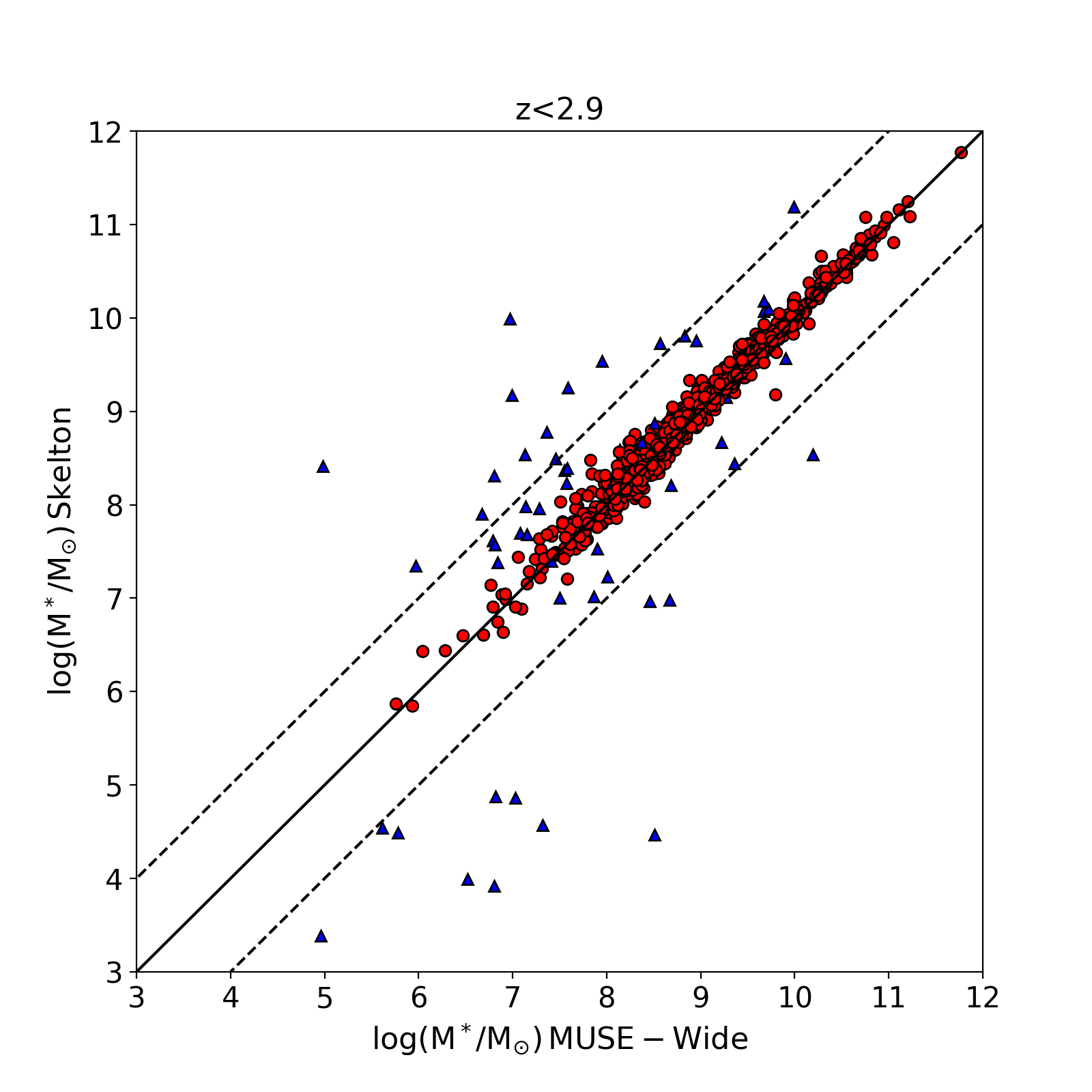}
\includegraphics[width=9.0cm]{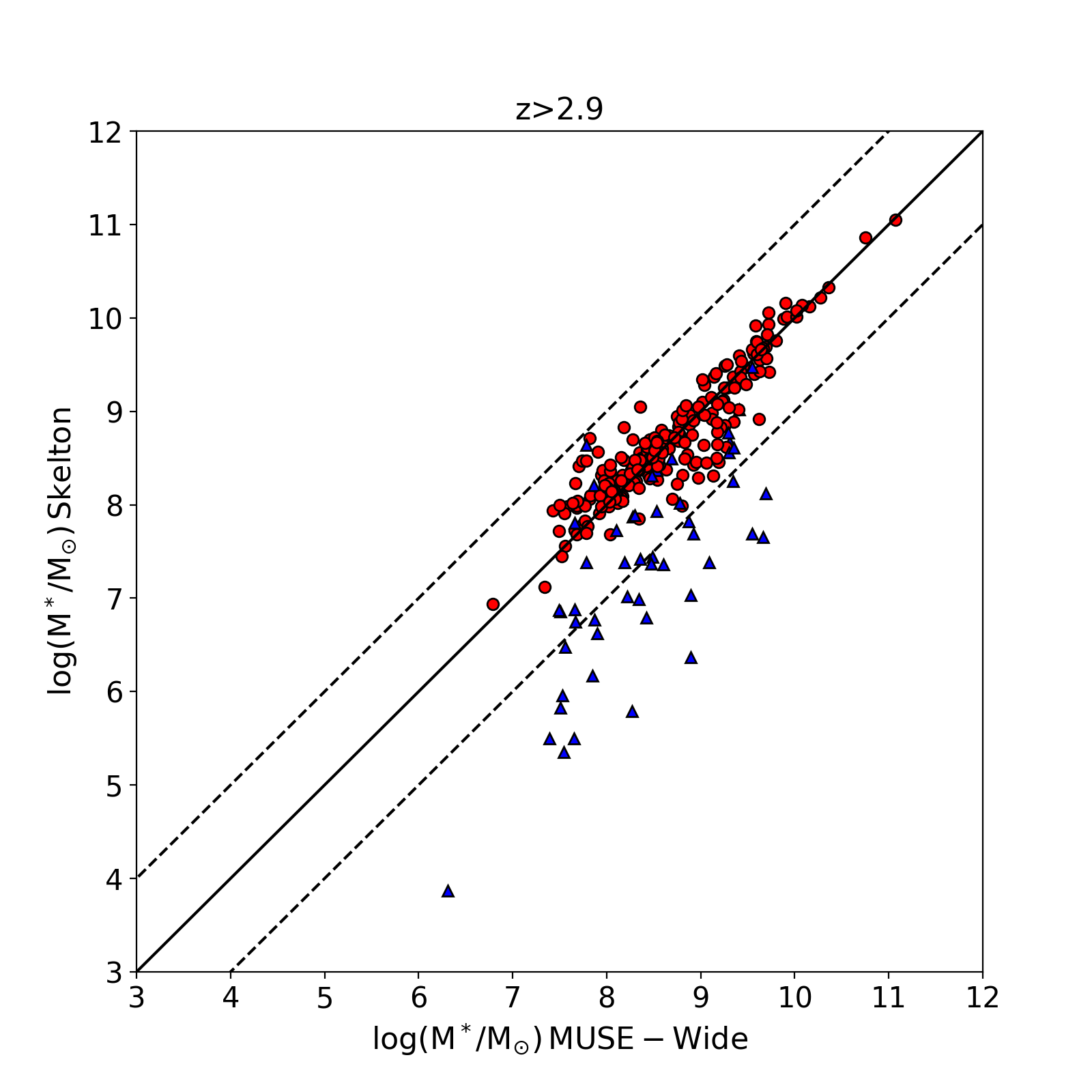}
\caption{Photometric mass estimates fixed at the MUSE-Wide
  spectroscopic redshifts compared to the \cite{skelton} photometric
  mass estimates based on photometric redshift estimates. The solid
  line represents the 1:1 mass equivalency, while the dashed lines
  show $\pm$1 dex differences between the two stellar mass
  estimates. Marked with blue triangles are catastrophic redshift
  outliers, which will be the main cause of discrepancies. At low
  redshift there is good agreement between the stellar masses, while
  at high redshift the scatter between the masses is larger due to the
  large photometric errors associated with these faint
  galaxies.}\label{masscomp} 
\end{figure*}

In Figure \ref{masscomp} we compare our derived stellar masses with
the ones from \cite{skelton}. The largest differences in the stellar
mass estimates come from catastrophic failures for the redshift
estimation from photometric data, so we display them with blue
triangles and do not take them into account for our statistics. As the
plot shows, the unlikely stellar masses of
log(M$^*$/M$_{\odot}$) $\ll 5.0$ in the Skelton data are due to the
galaxies being assigned an incorrect, much lower redshift than they
actually have. At low redshifts we find good agreement between the
stellar masses where there are no catastrophic redshift failures, with
a standard deviation of the differences between the masses of 0.11
dex. At high redshifts, the scatter is larger, here the standard
deviation of the differences between the stellar masses is 0.28 dex
with some stellar mass differences almost reaching 1.0 dex. This is
understandable as the photometry has larger errors for these fainter
sources, hence the probability distribution functions for the $\chi^2$
minimization will be broader. Furthermore, also here redshift errors
play a role, in particular the systematic shift of $\Delta z\sim0.2$
already noted can introduce an overestimate of $\sim$0.1-0.15 dex
\citep{nanayakkara16}. 

\subsection{Final emission line catalogs}\label{emcats}

Similar to \citetalias{herenz17}, we created two catalogs from our
emission line search. One is an object catalog in which the
information (redshift, confidence, photometric counterparts, etc.) for
the emission line galaxy is stored. Its UNIQUE\_ID is determined by
the leading emission line as described above. The other is an emission
line catalog, which is cross-referenced to the object catalog through
the UNIQUE\_ID. In it the physical properties of all emission lines
(coordinates, extent, flux, etc.) are listed, including all secondary
lines associated to an object. 

In addition to the 1D PSF weighted spectra (see Section
\ref{emdetect}), which are optimized to provide the best S/N in the
emission lines, we also extracted aperture spectra with the Kron
radius as the aperture radius (with a lower limit for the radius of
0.6\arcsec). While these spectra tend to be noisier, as they include
regions that are not strongly line-emitting, they capture the
spatially extended flux of the emitting galaxy withot a bias towards
the emission line region. Most of the emission line galaxies are more
extended than the PSF, hence the aperture spectra are more realistic
representations of the flux emanating from the emission line galaxy as
a whole. As in \citetalias{herenz17}, both PSF weighted and aperture
spectra are stored as FITS binary tables with columns for both air and
vacuum wavelengths.

The catalogs and spectra are available on the MUSE-Wide data release
webpage \url{http://www.musewide.aip.de} (see Appendix
\ref{web-catalogs}). The main emission-line source catalog includes
links to subpages, which include images centered at the emission line
position, a link to download the two 1D spectra described above, a
cross reference to the \cite{guo} cross-matched subpage (see Section
\ref{photocat}) and a link to download a $6\arcsec \times 6\arcsec$
mini 3D MUSE cube centered on the emitter position. The entries and
formats for the different catalogs are described in the ``Database''
tab of the MUSE-Wide webpage (see Appendix \ref{webdata}), but are
similar to the catalog entries of \citetalias{herenz17} except for the
addition of two columns, one for a MUSE-Wide field identifier and one
for the stellar mass of the object (described in Section
\ref{masses}).

\subsection{Cross-match with other multiwavelength catalogs}

\subsubsection{X-ray}

We then cross-matched our emission line objects with the X-ray source
catalog from the CDFS-7Ms {\it Chandra} observations
\citep{luo17}, the sources of which, are dominated by
AGN. Traditionally blind emission line surveys have unveiled numerous
active galaxies, so we expect some of the MUSE-Wide emission line
galaxies to contain AGN, too. At redshifts $z<0.4$, where \Ha\, and
other optical emission lines fall within the MUSE wavelength range, we
can employ the BPT diagram \citep{bpt} to distinguish AGN from
star-forming galaxies. At higher redshifts, however, we either need to
resort to other classification schemes \cite[which are however more
ambiguous, e.g.,][]{mex} or perform expensive near-IR spectroscopy to
obtain the rest-frame optical lines. X-ray data can help in
distinguishing the galaxies, particularly if the source is X-ray
luminous. Some weak X-ray sources can be driven by star-formation, but
their signal is of low luminosity and very soft, as it represents the
energetic tail of a thermal signal. 

\begin{table*}
\centering
\caption{Table of matched X-ray objects with only photo-$z$ values or
  disagreement between the redshifts}\label{x-table} 
\begin{tabular}{c c c c c c c}
\hline
ID & $z$ & ID & Separation & X-ray flux & $z$ & Redshift \\
MUSE-Wide & MUSE & {\it Chandra} 7Ms & (\arcsec) & (erg/s/cm$^2$) & 
{\it Chandra} & Source \\
\hline\hline
102028132 & 1.379 &  335 & 0.54 & 3.319e-17 & 1.038 & zSpec \\
105027078 & 0.681 &  364 & 0.85 & 8.972e-17 & 0.343 & zSpec \\
106036089 & 0.905 &  344 & 0.82 & 6.638e-17 & 0.956 & S14    \\
113001007 & 0.232 &  508 & 0.27 & 2.411e-17 & 0.220 & S14    \\
113022070 & 0.832 &  436 & 0.60 & 2.660e-17 & 0.854 & H14   \\
116003060 & 1.364 &  634 & 0.15 & 3.899e-17 & 1.363 & H14   \\  
117034085 & 0.228 &  693 & 1.11 & 6.435e-17 & 2.302 & zSpec \\
118011046 & 0.577 &  784 & 0.79 & 3.802e-17 & 0.270 & zSpec \\
123005089 & 0.544 &  640 & 0.49 & 4.219e-17 & 0.552 & H14   \\
123048186* & 4.379 &  654 & 1.40 & 2.462e-17 & 1.839 & H14   \\
123051191* & 4.507 &  625 & 1.35 & 2.102e-17 & 2.616 & H14   \\
124037072 & 1.003 &  710 & 0.97 & 6.136e-17 & 1.619 & zSpec \\
137003006 & 0.309 &  505 & 0.51 & 3.171e-17 & 0.308 & H14    \\
139073330 & 1.446 &  631 & 0.47 & 3.962e-17 & 1.511 & H14   \\
143041126 & 0.468 &  327 & 0.37 & 2.104e-17 & 0.456 & H14   \\
146002220 & 2.961 &  205 & 1.14 & 1.557e-16 & 1.610 & zSpec \\
\hline
\end{tabular}
\tablebib{S14 = \citet{skelton}; H14 = \citet{hsu14}}
\end{table*}

A cross-match was achieved when an emission line source is within 3
times the X-ray positional accuracy (SIGMAX in Table 4 of the
\cite{luo17} catalog). In \citetalias{herenz17} we required the X-ray
source to be luminous (have an ``AGN'' flag associated with it); we
drop that requirement as to classify even the faintest X-ray sources
as they may be at high redshift. One of the goals of extending the
X-ray imaging in this field to such long integration times was to find
intermediate and low luminosity AGN at high redshift for investigating
the faint end of the AGN luminosity function at $z>4$. This may
provide clues about the relative contribution of quasars to the
reionization of the Universe \citep{giallongo15}, and could also
constrain the earliest growth periods of black holes \citep{weigel15}.  

We match 127 emission line sources to X-ray counterparts after purging
2 superpositions from the X-ray catalog. As there have been extensive
campaigns to identify the X-ray sources since the CDFS began observing
in the early 2000s (e.g. \cite{szokoly04}; see \cite{luo17}, for the
26 literature references used for spectroscopic redshift
determination), it is not surprising that most of the sources have a
spectroscopic redshift associated with them. We are nevertheless able
to assign spectroscopic redshifts to sources which previously only
had photometric redshifts. All but 6 spectroscopic redshifts and 2
photometric redshifts are in excellent agreement with each other. In
some cases, when the separation between the sources is large, the
spectroscopic redshift may refer to another object and in other cases
there may be a misidentification of the spectrum. The discrepant cases
and the newly identified sources with only photometric redshifts are
listed in Table \ref{x-table}, marked S14/H14 for new identifications
(10) and zSpec for redshift discrepancies (6). We note that the two
high redshift MUSE-Wide sources marked with stars are not associated
to the optical counterpart for which the photometric redshift was
derived and due to their large distance to the X-ray source are likely
not associated with it either. 

Perhaps surprisingly we did not identify any new high redshift
($3<z<6.5$) sources in the X-ray population with our emission line
sample. The only two emission line sources above redshift of 3.0 are
well known Type 2 QSOs (MUSE-WIDE IDs: 104014050, 115003085), both of
which have the highest \Lya\, fluxes in our survey
\citep{norman02,mainieri05}. In the future we plan to match the X-ray
catalogs also to non-emission line sources and blindly extract
MUSE-spectra at the X-ray position to peer further into AGN
demographics at high redshift, but that is beyond the scope of this
paper.  

We may, however, characterize possible AGN emission from our high
redshift LAEs to determine whether there is low luminosity X-ray
activity coming from that population. Previous studies
\citep{treister13,vito16} have explored the possibility of identifying
early black hole growth by stacking several high redshift galaxies in
the CDFS. AGN at high redshift  are less affected by obscuration as
the high energy window moves into the {\it Chandra} spectral
range. While \cite{treister13} did not find any signal among luminous
LBGs, \cite{vito16} found a significant X-ray emission from massive
galaxies at $z \approx 4$, however attributing it to mainly star
formation processes. They speculate that since they did not find
dominant AGN features, either the dominant AGN population is in
fainter galaxies, the processes are hard to see or that most of black
hole growth occurs at later times in the Universe. This also implies a
flattening of the AGN X-ray luminosity function at high redshift.

Having a sample of 477 fainter high redshift sources at hand (479 LAEs
minus the two QSOs mentioned above), we also stacked the corresponding
CDFS X-ray data. We used
CSTACK\footnote{\url{http://lambic.astrosen.unam.mx/cstack_v4.32/}}
\citep{cstack} a web-based stacking tool, which takes into account the
intricacies of exposure maps, response matrices, PSF variations,
etc., for various deep {\it Chandra} fields. At the time of the
analysis the 7Ms data \citep{luo17} was not yet freely available on
the CSTACK interface, so we use the 4Ms data \citep{xue11} for
stacking, yet a subsequent stacking analysis done by scientists with
access to the 7Ms data yielded similar results (Vito, private
communication). 

\begin{figure}
\centering
\includegraphics[width=3.7cm]{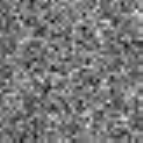}
\hspace*{1.3cm}
\includegraphics[width=3.7cm]{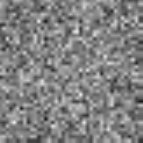}
\includegraphics[width=4.3cm]{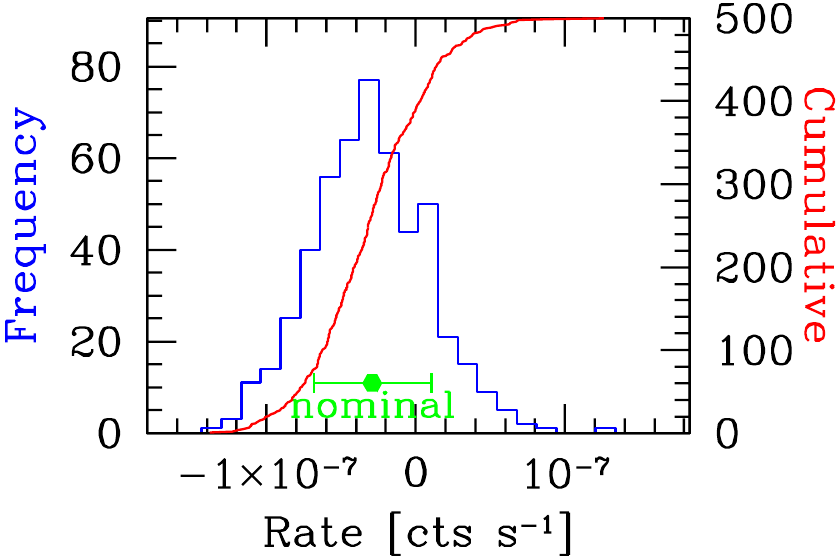}
\hspace*{0.1cm}
\includegraphics[width=4.3cm]{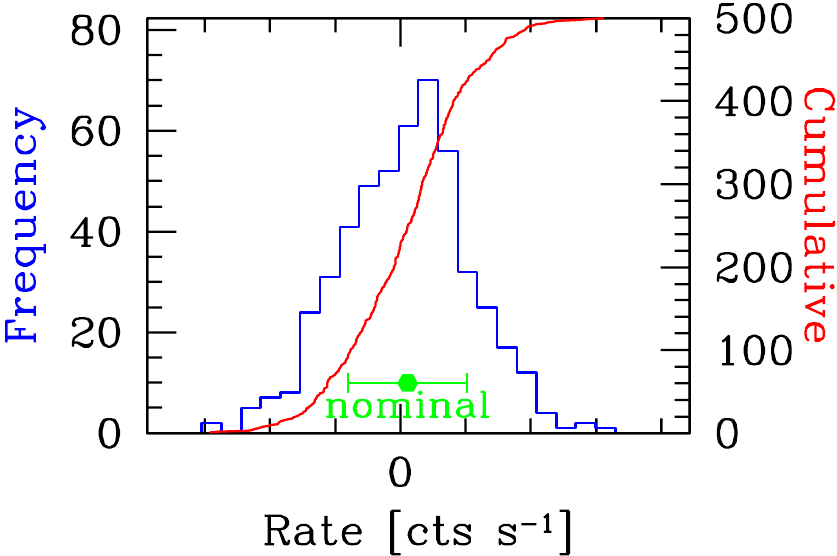}
\caption{Null result of stacking X-ray images from the 4Ms CDFS data
  at LAE positions. The panels on the left refer to the soft band
  (0.3-2.0keV), the ones on the right to the hard band
  (2.0-7.0keV). The upper panels show the stacked images, whichare
  20\arcsec $\times$ 20\arcsec interpolated on a 0.5\arcsec pixel size
  Chandra grid. The lower panels describe the statistics of the mean
  stack.}\label{xstack}  
\end{figure}

Figure \ref{xstack} shows the results of the X-ray stacking. The mean
count-rate for both the low and high energy band is consistent with
zero. The combined stack has a 3$\sigma$ upper limit of
$2.8\times10^{-8}$ cts/s or $3.6\times10^{-18}$ erg/cm$^2$/s in X-ray
flux\footnote{Using PIMMS:
  \url{http://cxc.harvard.edu/toolkit/pimms.jsp} and an unabsorbed
  power law model at redshift z=4.0 with Galactic N$_H=
  8.8\times10^{19}$ cm$^{-2}$. We employed the {\it Chandra} Cycle 10
  ACIS-I filter for the countrate to flux conversion}. The population
of LAEs does not show any AGN activity in the stacks and is therefore
not dominated by it, at least not by typical accretion processes,
which produce an X-ray corona. Either black hole growth is a
stochastic process occurring only in a small fraction of galaxies in
the early Universe leaving the bulk of the black hole growth to happen
at a later time or the processes are radiatively inefficient and/or heavily
obscured, e.g. super-Eddington accreting Broad Absorption-Line Quasars
which are X-ray weak \citep{luo14}.

\subsubsection{Submillimeter and Radio}

\begin{figure}
\centering
\includegraphics[width=9.2cm]{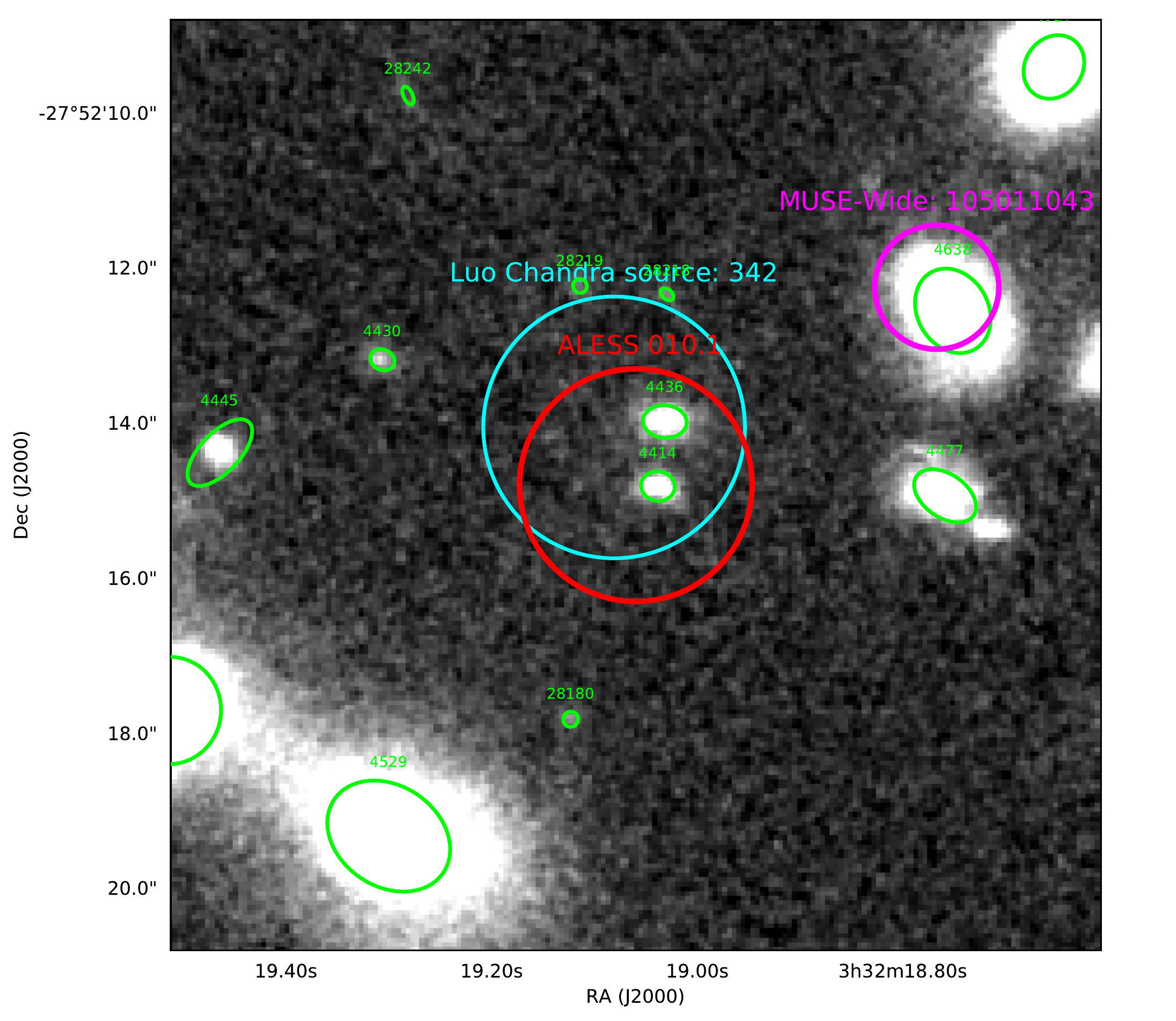}
\caption{CANDELS WFC3 $6\arcsec \times 6\arcsec$ cutout of the region
  centered on ALESS 10.1. CANDELS photometric objects from \cite{guo}
  are marked in green, a MUSE-Wide emission galaxy with a $0.8\arcsec$
  radius in magenta, the X-ray source with its corresponding error
  radius in cyan and the ALESS source with an arbitrary $1.5\arcsec$
  radius in red.}\label{aless10} 
\end{figure}

An interesting question is whether LAEs can be related to
submillimeter galaxies, which are also forming stars at a high rate or
host AGN, but usually belong to a far dustier population of
high-redshift sources. Unfortunately, of all the continuum
submillimeter sources catalogued with ALMA in the CDFS in the ALESS
survey \citep{alma-cdfs} only one falls within the fields of view of
MUSE-Wide DR1. ALESS 10.1 is in candels-cdfs-05, is centered on
CANDELS \# 4414 and also matches the X-ray source 342 of
\cite{luo17} (see Figure \ref{aless10}). There is no match to our
emission line catalog (MUSE-Wide 105011043 is nearby, but is clearly
attributed to another galaxy). The source shows has a steeply rising
SED all the way out to the mid-IR without any noticeable breaks from
the broad band filter fluxes. \cite{hsu14} assign a photo-$z$ of $z \sim
0.76$ to the source, but we cannot confirm this from our optimally
extracted spectrum of that source (see Section \ref{tdose}) as it
shows no identifiable spectral features. 

\begin{table*}
\centering
\caption{Emission line sources match to VLA radio catalog}\label{radiotable}
\begin{tabular}{c c c c c c c c}
\hline
ID & z & ID & Separation & S/N & flux density & X-ray & Correct \\
MUSE-Wide & MUSE & VLA & (\arcsec) & radio & 1.4GHz ($\mu$Jy) & counterpart? & crossmatch? \\
\hline\hline
105002016 & 0.343 & 117 & 0.42 &  7.2 & 106 & yes & yes \\
120023032 & 1.119 & 192 & 0.45 & 10.3 & 172 & yes & yes \\
122025120 & 0.670 & 150 & 0.81 &  4.4 &  42 &  no & yes \\
123028137 & 0.545 & 157 & 0.71 &  4.9 &  48 & yes & yes \\
124002008 & 0.242 & 163 & 0.29 &  6.6 &  63 & yes & yes \\
136002114 & 0.247 & 136 & 0.50 &  9.4 &  86 & yes & yes \\
136010134 & 2.224 & 145 & 0.55 &  9.7 & 129 & yes & yes \\
136034182 & 0.736 & 139 & 0.89 & 26.0 & 265 & yes & yes \\
143021059 & 0.734 & 112 & 0.67 & 59.1 & 524 & yes & yes \\
146002220 & 2.961 &  91 & 0.87 & 17.9 & 164 & yes &  no\tablefootmark{a}\\
146026263 & 0.576 &  95 & 0.76 & 20.9 & 201 & yes & yes \\
\hline
\end{tabular}
\tablefoot{
\tablefoottext{a}{Already noted as mismatch in Table \ref{x-table},
  see text for details.}
}
\end{table*}

As with X-ray sources, radio surveys are ideally matched to emission
line surveys, because the physical conditions that produce radio
emission (synchrotron emission for AGN, thermal free-free emission
from star-forming HII regions) also may produce emission lines. We
cross-match the 464 sources from the Very Large Array (Jansky Array,
VLA) 1.4GHz survey over the extended-CDFS \citep{vla-cdfs,miller08} to
our emission line sources. As radio positional errors were small
($\lesssim 0.2\arcsec$), we used a fixed $1.0\arcsec$ as our matching
radius. 11 radio sources mathc and all but one of the matches to the
emission line catalog are also X-ray sources listed in the
\cite{luo17} catalog. All have previous spectroscopic identifications
in the Literature and all but one agree with the MUSE-Wide redshifts
(see Table \ref{radiotable}). The redshift mismatch belongs to
MUSE-Wide source 146002220, which has only confidence 1, its S/N is
very close to the limit of 5.0 and it may not be associated to the
correct source (see also Table \ref{x-table}). The much closer source
to the X-ray and radio position is CANDELS ID \#16375. It does not
fall in our emission line catalog and we are only able to identify it
as a $z\sim0.442$ galaxy with confidence 1 (see Section \ref{photid}).    

\section{Spectroscopy and identification of photometrically
  selected objects}\label{photocat}

One of the advantages of the concept of ``spectra of everything'' is
that we can actually extract a spectrum ``of everything'' even if at
the end it is too noisy for identification. A MUSE extraction showing
a noisy spectrum for a specific photometric object at least tells us
that the object does not have strong emission lines or other spectral
features, so that any future follow-up will require substantially
higher integration times. We decided to first concentrate on the
CANDELS \cite{guo} catalog for spectral extractions. In the next data
release also the 3D-HST \cite{skelton} catalogs will be used. 

There are 9,205 CANDELS objects within the MUSE-Wide DR1 survey
area. At the edges, we required the object position to have at least
two MUSE-Wide 15-minute exposures to be included in the catalog. We do
not create separate MUSE IDs for these photometrically selected
objects, rather we further use the CANDELS ID. Since we cross-linked
the CANDELS \cite{guo} catalog already in the emission line catalog,
the link back here was easy to implement. However, before we optimally
extracted the spectra, we needed to decide what to do for objects that
lie in the overlap regions and could therefore be extracted from up to
4 fields overlapping in those regions. To make an object always be
associated with a field in the survey we creatde a field map to return
the field number, i.e. candels-cdfs-37 would return 37. We describe in
the next section how we constructed such a field map.

\subsection{Field Map}\label{mapping}

Using the individual exposure maps, where each pixel on the map has a
value between 0 and 4, we created a merged exposure map, from which we
could attribute each exposure pixel to a field number. To create the
field map we assigned a field number to each pixel with an exposure
using the following procedure: (i) pixels with exposures
from only one field got that field number; (ii) pixels in which there
were more than one exposures we chose the field number with the
largest number of exposures; (iii) pixels in which there were the same
number of exposures for the fields, we chose the field according to a
priority table relating to data quality. The priority table was sorted
by an estimate of the noise properties in the cube, basically the
amount of sky affecting the spectrum and therefore increasing the
noise in those spectra. We did this by sorting by the inverse of a
product of sky emissivity and the square of the seeing FWHM (Gaussian
$p_0$) of each field. Fields 21, 22, 17 therefore had the highest
priorities, while fields 40, 42, 26 had the lowest. Finally, we got
rid of non-contiguous, non-connected, ``island pixels'' that may occur
at the edges, especially at the slicer stack transitions, so to have a
smooth field preference over the map.  

\begin{figure*}
\centering
\includegraphics[width=14cm]{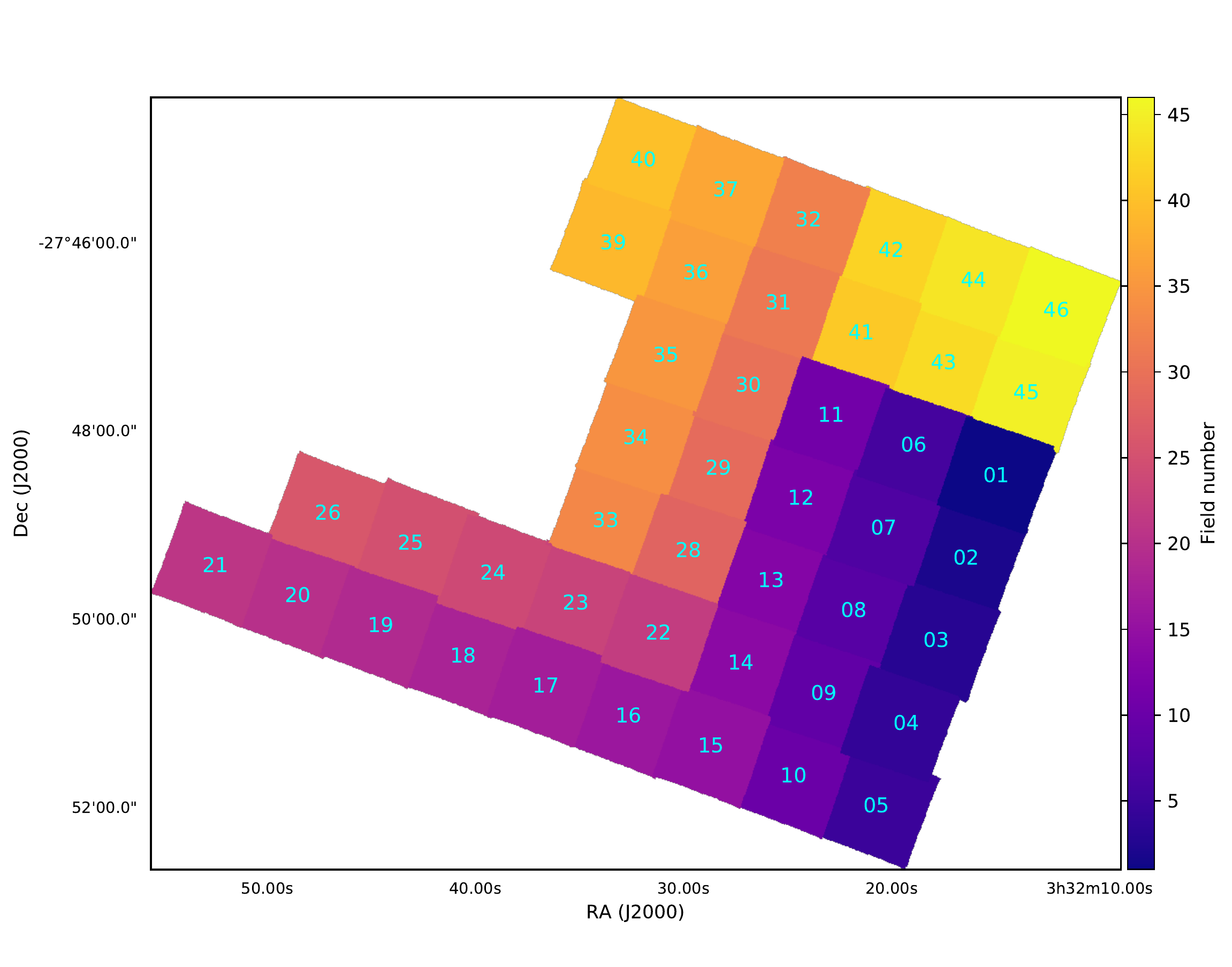}
\caption{MUSE-Wide field mapping. It returns a preferred field to use
  for any location located within the MUSE-Wide footprint based on the
  amount of exposures and the best observing conditions available for
  each coordinate.}\label{fieldmap}
\end{figure*}

An image of the field map is shown in Figure \ref{fieldmap}. One sees
that fields with mediocre observing conditions, surrounded by less
noisy fields such as candels-cdfs-29, have overall less space in the
survey assigned to them. Each object of the photometric catalog got
assigned to exactly one MUSE-Wide field based on this field mapping.

\subsection{Optimal 3D spectral extraction}\label{tdose}

One further step was to extract 1D spectra of each individual source
in our MUSE-Wide photometric catalog from the MUSE data cubes. To
optimally extract flux-conserving spectra from 3D data, i.e. spectra
with the best possible signal-to-noise ratio, the spatial morphology
and the wavelength dependent PSF described in Section
\ref{psf-estimation} needed to be accounted for. Furthermore, good
deblending of neighboring sources was essential to produce spectra
representing the intrinsic spectra as accurately as possible. The
spectra provided with the current data release and the classification
of the brightest sources described in the next Section were extracted
using version 2.0 of the dedicated tool ``Three Dimensional Optimal
Spectral Extraction'' (TDOSE). TDOSE generalizes the concepts of
\cite{pampelmuse} to extended non-PSF sources, i.e. galaxies. TDOSE
will be presented in Schmidt et al. in prep, and is publicly available
from GitHub\footnote{\url{https://github.com/kasperschmidt/TDOSE}}. In
the following we provide a short summary of the spectral extraction
performed for this study but refer the reader to the above sources for
further details on the tools and methods involved in the spectral
extraction.

TDOSE uses PSF-convolved morphological models of photometric
counterparts to {\it simultaneously} extract 1D spectra from the 3D
MUSE cubes. This is done by solving a set of linear equations
minimizing the difference between the object models convolved with the
wavelength dependent MUSE PSF cube, scaled by some flux, and the
observed flux in the final MUSE datacubes described in Section
\ref{effectivenoise}. To make this process fully analytic we used the
Gaussian PSF estimation and described the galaxies by single
multi-variate Gaussian models \citep{hogg13} or point sources. The
latter models were used for objects with detected emission lines in
the MUSE data (Section \ref{emline}), but no or minimal photometric
counterpart. The morphological models were generated using
the\texttt{curve\_fit} optimizer of
Scipy\footnote{\url{https://www.scipy.org}} optimizer on the F814W
CANDELS {\it HST} images. The F814W images were chosen due to their
depth as well as the peak sensitivity wavelength and similarity in
wavelength coverage of this filter ($\approx 6880$ -- 9640~\AA) with
the MUSE wavelength coverage (4750 -- 9350~\AA).  

Using single multi-variate Gaussians as source models, is obviously
simplifying the often complicated morphology of galaxies. However,
after convolving the {\it HST}-based models with the MUSE PSF, the
loss of flux between the spectral extractions of single-component
Gaussian model and a GALFIT \citep{galfit} multiple-component
\cite{sersic} model is only $\sim$5\% at the \OII\, wavelength for a
sample of $\sim$150 relatively bright ($23 < \mathrm{F814W} < 24$)
\OII\, emitters in the MUSE-Wide CDF-S footprint. The exact amount of
flux lost from using a simplified single-component model, of course
depends on the intrinsic ({\it HST}) morphology of the sample studied. 

Solving the set of linear equations simultaneously ensured optimal
deblending of sources, when assembling the 1D spectra. TDOSE
essentially assigns a fraction of the flux in each voxel of the MUSE
datacube to objects contributing to these voxels according to their
morphological PSF-convolved models. Solving the system of equations
provided wavelength dependent flux scalings for each individual source
(the sum of the factional flux scalings for all voxels) in the field-of-view of
the MUSE datacubes. These $\chi^2$ minimizing flux-scalings were the
1D spectra used for object classification and are available on the
data release webpage \url{https://musewide.aip.de} (see Appendix
\ref{webpage}). 

The main assumption of the spectra extracted as part of
this study, is that the light distribution is a multivariate Gaussian
and follows the F814W continuum light distribution convolved with the
Gaussian MUSE PSF at each wavelength of the MUSE datacubes. Given that
most galaxies are known to follow non-Gaussian light distributions,
e.g. \cite{sersic} profiles, or have multiple components and distinct
features that might very well vary with wavelength, these assumptions
are be somewhat restrictive and may depend the scientific applications
of the spectra, but are only expected to results in flux-losses on the
few-percent level, as described above. More fundamental, features that
do not intrinsically follow the light-distribution of the continuum
are expected to be biased in the TDOSE spectra.. In particular, as
\Lya\, emission is known to be more extended than the continuum
emission \citep{wisotzki16,leclercq17}, the TDOSE spectra are
unsuitable for estimating total \Lya\, fluxes for the brightest LAEs
in the MUSE-Wide sample. The same holds for the study of any other
spectral features that do not, to a fair approximation, follow the
distribution of the continuum light. This presumably includes all
nebular emission lines, which often do not trace the spatial
distribution of the stars that make up the continuum light
modeled. Hence, the TDOSE spectra are "optimal" for the continuum
sources, but are not necessarily so for wavelengths with emission lines. 

However, for the public data release and for the classification of the
brightest photometrically selected objects in MUSE-Wide based on their
spectral features, the Gaussian assumptions are sufficiently
detailed. We furthermore chose to use these simplifying assumptions,
as they made the spectral extraction process fully analytic and
provided a homogeneous sample of spectra for all objects in the data
release, irrespective of object type, magnitude, size or redshift. 

\subsection{Identification of galaxies brighter than 24th
  magnitude}\label{photid}

Of the 9,205 photometrically selected objects, 772 have a F775W
magnitude brighter than 24. We inspect the objects in this subset of
the photometric catalog further to identify them
spectroscopically. Even though we use MUSE spectra to classify this
bright subset of the photometric catalog, we do not create separate
MUSE IDs for the objects in this catalog, but keep the CANDELS ID for
our classification. Since the photometric catalog links to the
emission-line objects, we can discard objects, that have already
assigned a redshift. 499 (64\%) already were identified via their
emission lines, we adopt the identification gleaned from the emission
lines, leaving us to inspect 273 objects using the optimally extracted
1D TDOSE spectra. These non-emission line sources were primarily
identified through their absorption features.  

\begin{figure}[ht!]
\centering
\includegraphics[width=8.8cm]{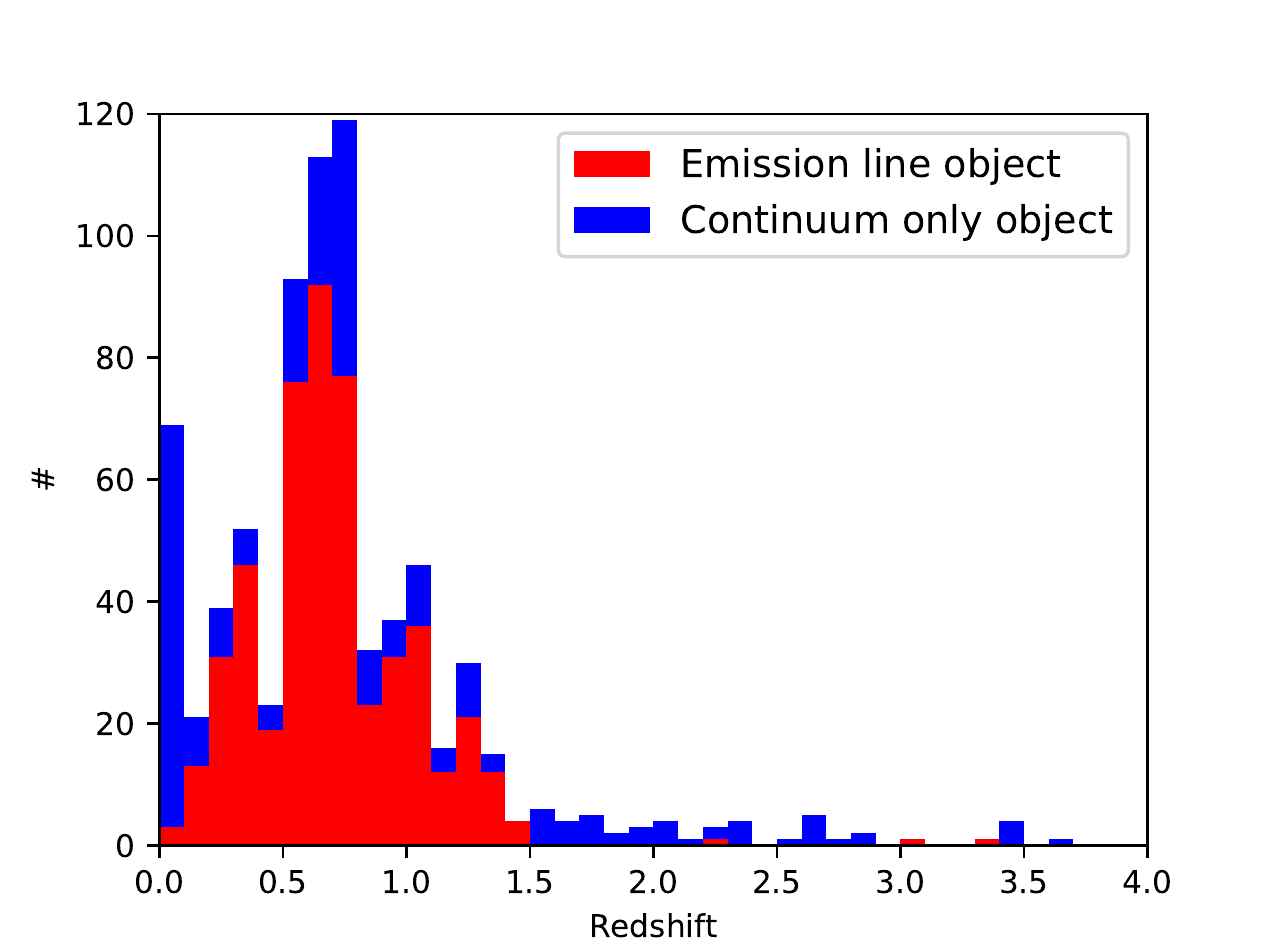}
\includegraphics[width=8.8cm]{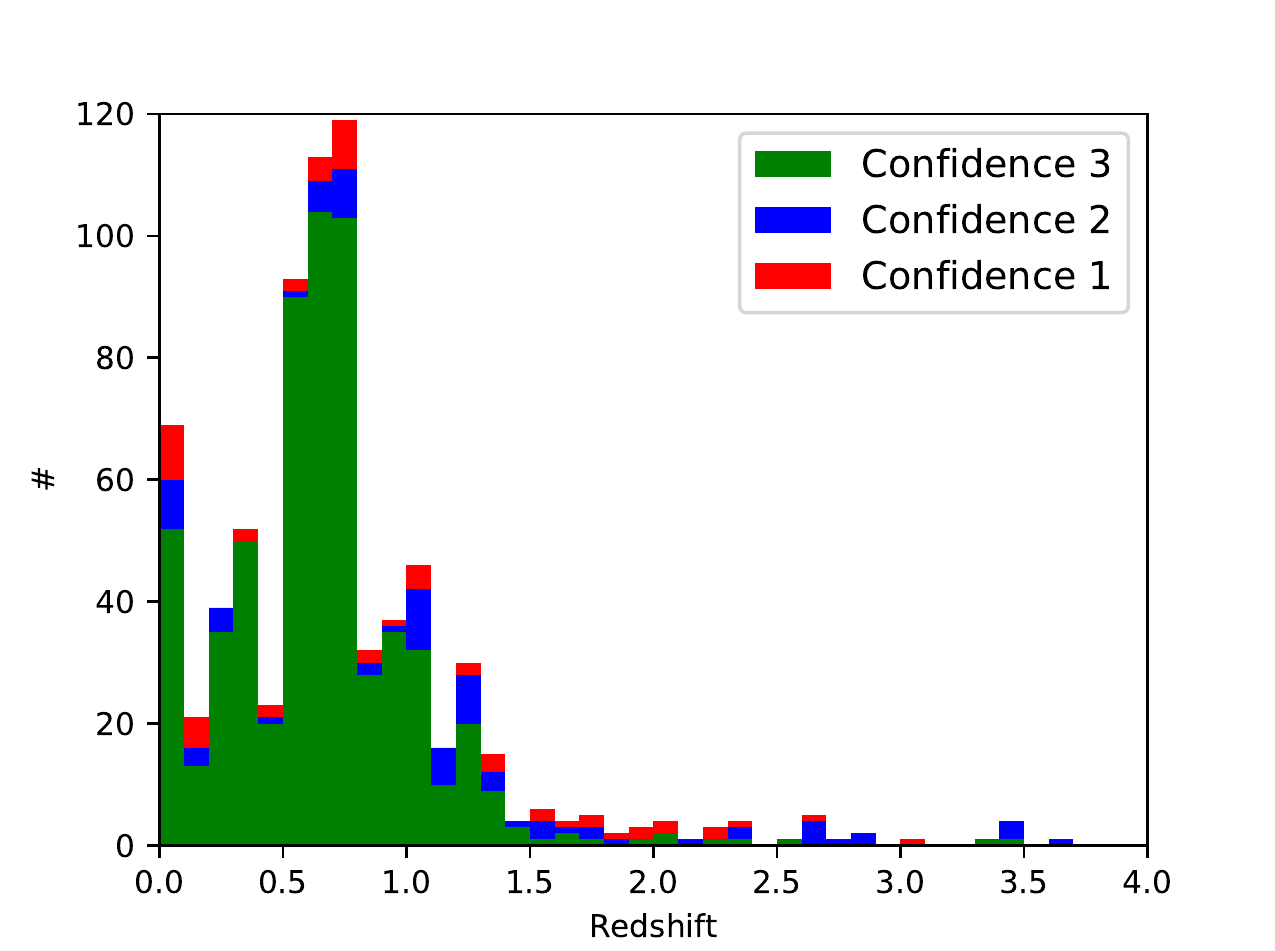}
\caption{Redshift identification of photometrically selected objects
  brighter than 24th magnitude. Shown are stacked histograms of the
  redshift distribution. The top panel shows the identification by
  type of object (emission-line object or continuum only object),
  while the lower panel shows the distribution of the confidence
  levels of the identification.}\label{redshift-photo}
\end{figure}

Similar to the MUSE-Deep approach \citep{inami17}, we used a
modified version of the redshift fitting software MARZ
\citep{marz}. MARZ determines redshifts based on a version of the
AUTOZ cross-correlation algorithm \citep{autoz}. We employed the
redshift templates also used for MUSE-Deep. The cross-correlation was
done with our TDOSE spectra by a team of investigators (L.W., K.B.S.,
D.K. and T.U.) first individually and later consolidated by agreeing on
a template and a confidence number. Also here the confidence levels
are somewhat subjective, but we tried to anchor it to the same rate as
for the emission line sources, i.e. expecting less than 1\%, 10\% and
50\% false identifications for confidence 3, confidence 2 and
confidence 1 objects, respectively. We were able to identify 98\% of
the 772 objects with only 15 objects (2\%) remaining unidentified. We
release this identification subcatalog on the photometric catalog page
of our data release page (see Appendix \ref{web-catalogs}). 

Figure \ref{redshift-photo} shows the redshift distribution for the
757 identified objects brighter than 24th magnitude (772 objects - 15
non-identifications). Since this is a magnitude-limited sample, it is
expected that the redshift distribution skews towards low redshift
objects. Nevertheless we are able to identify some objects in the
``redshift-desert'' via their {\mbox{Mg\,\sc {ii}}}, {\mbox{Fe\,\sc
    {ii}}} and {\mbox{Al\,\sc {iii}}}\, absorption features. Our imposed 24th
magnitude limit is clearly a conservative choice that ensures high
success rate. Probably there are several objects fainter than 24th
magnitude that are not in the emission line catalog, but that could be
nonetheless be identified spectroscopically either through faint or
broad emission lines not captured by LSDCat or through absorption
lines associated with strong SED features (e.g. Ca H+K absorption with
the 4000\AA\, jump). This is beyond the scope of this paper.

The photometric catalog and the identification catalog for objects
brighter than 24th magnitude are stored in the MUSE-Wide webpage (see
Appendix \ref{webpage}). The description of each field is given in the
``Database'' tab (see Appendix \ref{webdata}). We have tried to keep
the columns as simple and self-explanatory as possible providing only
a coordinate, a few {\it HST} magnitudes and a possible association to
the emission-line catalog for each object. The identification catalog
only has two additional columns, one for the redshift, one for the
confidence level of the identification.

\section{Conclusions and Outlook}

We have presented the first data release of the MUSE-Wide survey, a
blind 3D survey targeting well-known and studied deep fields with
extensive multiwavelength data, such as the GOODS-S/CDFS and
CANDELS-COSMOS areas. DR1 encompasses 39.5 arcmin$^2$ over 44 MUSE
fields in the GOODS-S/CDFS observed at a depth of 1 hour. It
represents the wide, ``shallow'' component to the MUSE-Deep survey
carried out in the HUDF \citep{muse-deep}. 

\begin{figure*}
\centering
\includegraphics[width=15cm]{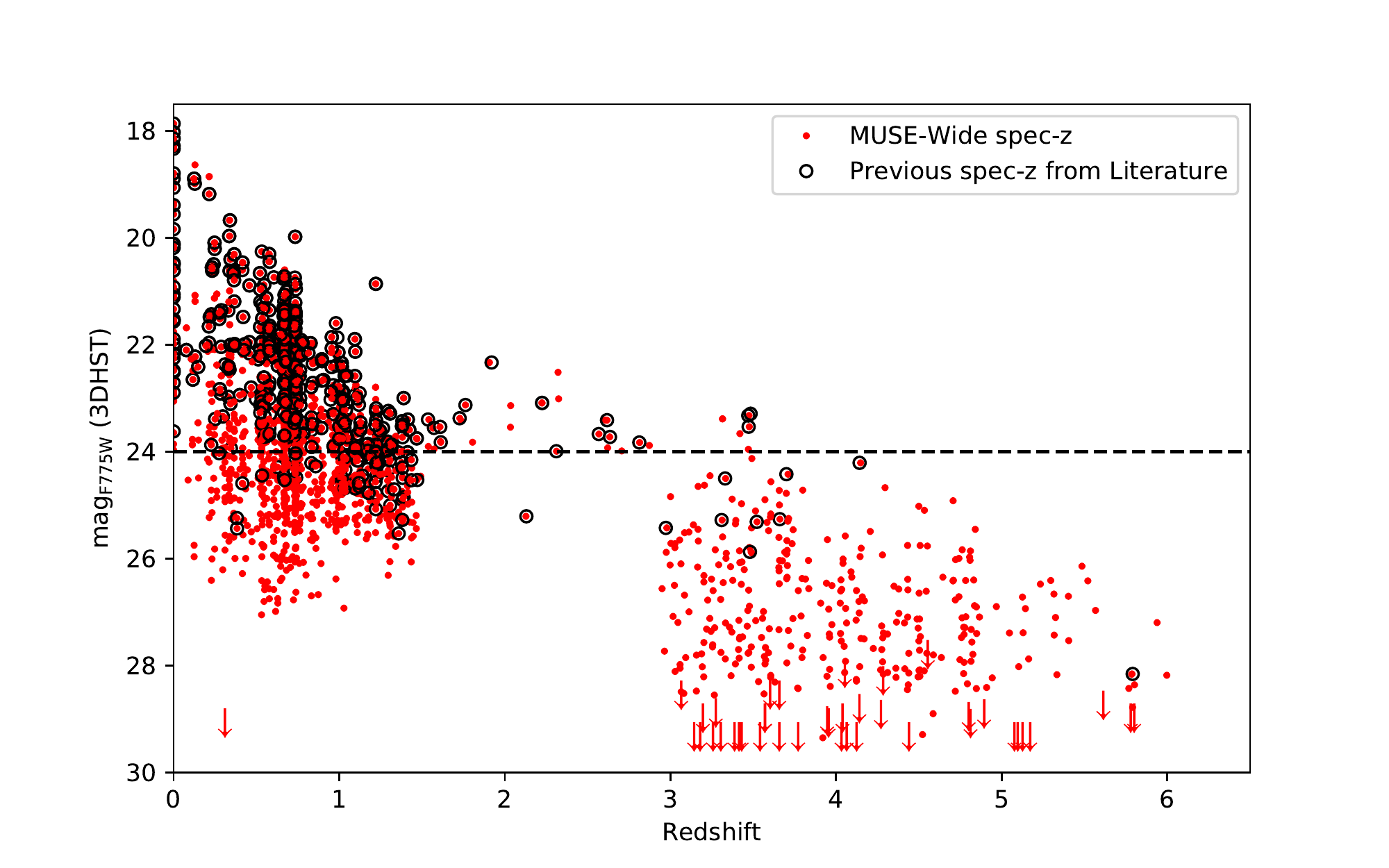}
\caption{F775W {\it HST} magnitude of both photometrically and
  emission-line selected high confidence sources (1,597 objects with
  confidence $\ge$ 2) as a function of redshift. The red dots and
  arrows (upper magnitude limits) represent the MUSE spectroscopically
  determined redshifts, while the open black circles denote previous
  spectroscopic redshifts of those sources from the
  literature. The large amount of redshift identifications, especially
  at high redshift highlight the capabilities of MUSE-Wide as an
  incredibly effective redshift survey.}\label{mag-redshift}   
\end{figure*}

The main scientific contributions in this data release (and of deep
blind 3D MUSE surveys in general) are the detection of emission line
sources in the datacubes and the optimal extraction of 1D spectra
based on the prior {\it HST} photometric information. Analysis of the
spectra of these emission line objects and bright photometric objects
yielded 1,859 spectroscopic redshift identifications (1,602 emission
line objects + 257 continuum only objects), with 1,597 of those being
of high confidence (confidence $\ge$ 2). Even with only one hour
integration times, this represents an unprecedented density of
spectroscopic redshifts gathered in extragalactic surveys, surpassing
even the deepest spectroscopic surveys by an order of magnitude
(e.g. LBG-z3 -- \citealt{steidel03}, FSF -- \citealt{fsf}, VUDS --
\citealt{lefevre15}, VANDELS -- \citealt{vandels}). 

Similar to Figure 18 in \cite{inami17} of the MUSE-Deep catalog, in
Figure \ref{mag-redshift} we show the F775W magnitude over redshift of
our high confidence sources. The black circles denote the sources with
previous spectroscopic redshift identifications. From this Figure it
is clear that MUSE-Wide opens up a new window at high redshift based
on the detection of the \Lya\, line at $z > 2.9$. Naturally, there are
some differences between the typical MUSE-Deep and MUSE-Wide
identifications, since the MUSE-Deep data extends well over a
magnitude deeper to analyze continuum spectra. We see this especially
pronounced in the ``redshift-desert'' at $1.5 < z < 2.9$, where
MUSE-Deep is able to identify a number of objects based on their
\CIII\, emission or the absorption features seen in that
range. MUSE-Deep offers an excellent opportunity to study large number
of LAEs, the densities ever increasing at the faintest
magnitudes. However, MUSE-Wide covers the LAEs with relatively bright
UV counterparts, which offers the opportunity to study their stellar
content with upcoming instruments, such as the James Webb Space
Telescope ({\it JWST}). Another area where MUSE-Wide can probe new
ground is in the identification of $z < 1.5$ emission-line dwarf field
galaxies with magnitudes $> 24$. Those sources are not typically
targeted in redshifts surveys, either because they are too faint or
because they have (correct) low photo-$z$ values. MUSE-Deep can also
easily find and identify them, however the density of these systems
does not increase as rapidly with magnitude as for the LAEs; the
spatial density seems to be better suited for a survey such as
MUSE-Wide, with the bulk of the population having magnitudes
mag$_{\mathrm{F775W}}$ = 24 -- 27. 

Of course, a 3D survey such as MUSE-Wide can act as more than a simple
redshift identification survey. The multiplexing capabilities of MUSE
let us further characterize the identified sources, such as
metallicity gradients at low redshift \citep{carton18} or \Lya-halos
at high redshift (Saust et al. in prep.).

A future and final data release will roughly double the current DR1 in
size covering all 91+9 MUSE-Wide fields, including the 23 fields in
the CANDELS-COSMOS region. We plan on improving the data reduction
using learned insights and new software tools that were released since
we froze our reduction pipeline early in the survey. The insights from
the first 44 fields already predict that we will succeed reaching our
goal of finding more than 1000 LAEs over the MUSE-Wide survey area.

\begin{acknowledgements}

The authors give thanks to the staff at ESO for extensive support
during the visitor-mode campaigns at Paranal Observatory.

We thank the eScience group at AIP for help with the functionality of the
MUSE-Wide data release webpage and in particular Ole Streicher for the
development of the cut-out tool. We thank Nicolas Bouch\'{e} for helpful
comments that improved the manuscript. 

T.U., L.W., J.K., K.B.S., E.C.H, R.S., C.D. and J.C. acknowledge
funding by the Competitive Fund of the Leibniz Association through
grants SAW-2013-AIP-4 and SAW-2015-AIP-2. R.B. acknowledges
support from the ERC advanced grant 339659-MUSICOS and the FOGHAR
Project with ANR Grant ANR-13-BS05-0010. P.M.W. acknowledges support
through BMBF Verbundforschung (project MUSE-AO, grant 05A14BAC).

This work has made use of the Chandra Deep Field South Dataset,
Dataset Identifier: ADS/Sa.CXO\#Contrib/ChandraDeepFieldS 

This work has made use of observations taken by the CANDELS
Multi-Cycle Treasury Program with the NASA/ESA HST, which is operated
by the Association of Universities for Research in Astronomy, Inc.,
under NASA contract NAS5-26555. 

Based on data obtained with the European Southern Observatory Very
Large Telescope, Paranal, Chile, under Large Program 185.A-0791, and
made available by the VUDS team at the CESAM data center, Laboratoire
d'Astrophysique de Marseille, France. 

This work has made use of observations taken by the 3D-HST Treasury
Program (GO 12177 and 12328) with the NASA/ESA HST, which is operated
by the Association of Universities for Research in Astronomy, Inc.,
under NASA contract NAS5-26555.

This research made use of Astropy, a community-developed core Python
package for Astronomy \citep{astropy}. 

\end{acknowledgements}

\bibliographystyle{aa} 
\bibliography{musewide-dr1.bib} 

\appendix

\section{Data Release Webpage}\label{webpage}

We have described the first public data release from the MUSE-Wide
survey, dubbed DR1. All data are available under the URL:
\url{https://musewide.aip.de}. The data release web page was created
using the Open Source Daiquiri
framework\footnote{https://github.com/aipescience/django-daiquiri}. DR1
contains the cubes, spectra and catalogs for the first 44 fields in
the candels-cdfs region. The page also functions as survey webpage
providing news and notifications about the MUSE-Wide survey. 

\begin{figure*}
\centering
\includegraphics[width=16.5cm]{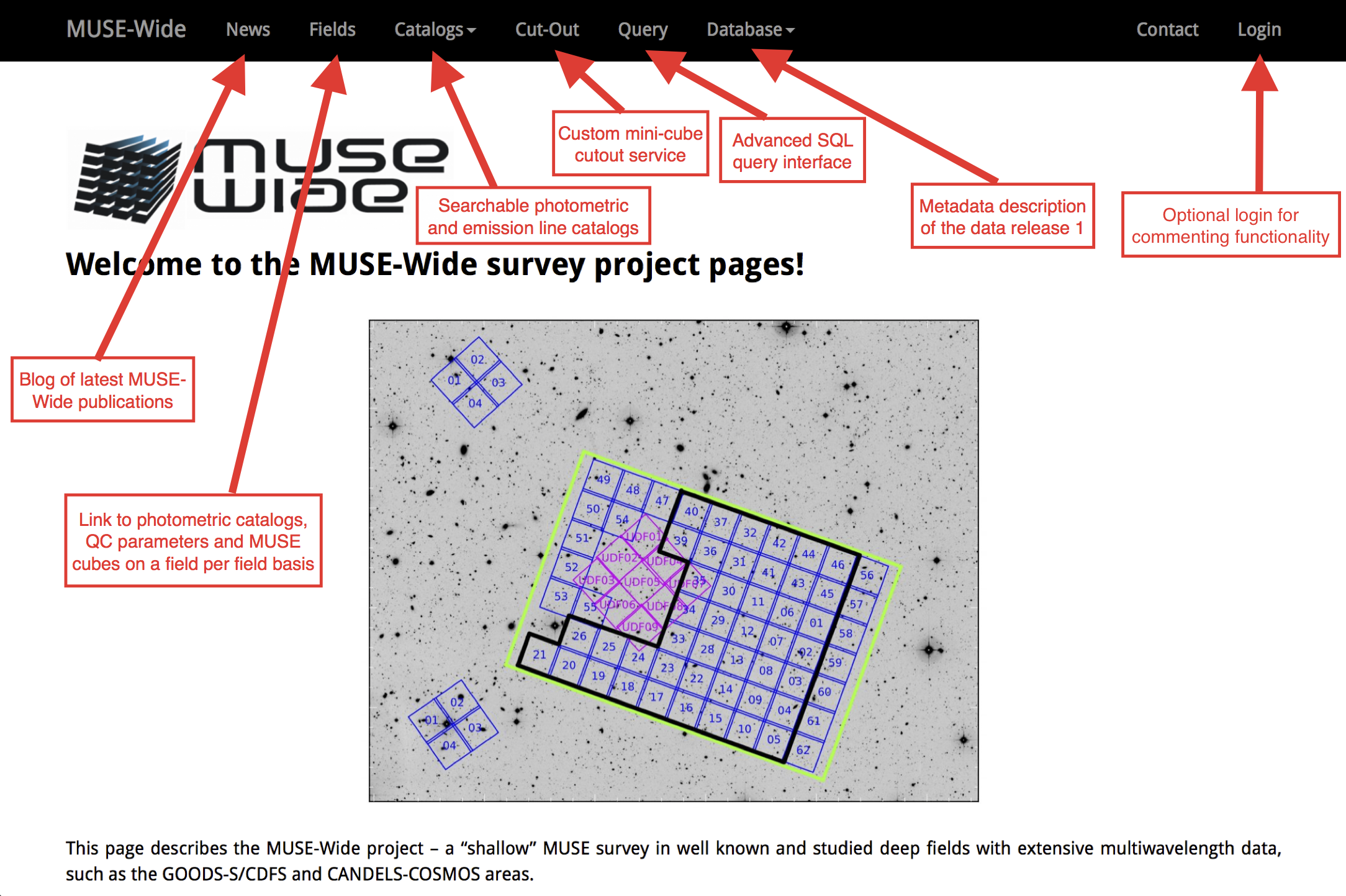}
\caption{Screenshot of the MUSE-Wide data release landing page. The
  data is accessed throughout the various tabs described in this
  Section.}\label{screenshot} 
\end{figure*}

A screenshot of the landing page as of October 2018 is shown in Figure
\ref{screenshot}. Most of the data is accessible via tabs. Clicking on
each tab opens up a new window, the contents of which are described in
the subsections below. The login is optional for blog notification,
commenting functionality and for storing SQL queries. Further data
releases will also occur via this webpage. Tables within the webpage
are registered with the Virtual Observatory, and each MUSE-Wide cube
and MUSE-Wide object (either emission-line source or TDOSE extracted
spectrum) has a citable DOI.  

\subsection{MUSE-Wide Fields}

The ``Fields'' page first lists links to our DR1 exposure map and the
field mapping to assign a field number to each coordinate within the
MUSE-Wide footprint. It then shows a table with 5 columns: the field
name, linking to an individual field subpage, the RA/Dec of the
center of the field, the datacube headers for each field and a link to
download the corresponding datacube. We warn the user that each cube
is about 5GB in size, which should be kept in mind when downloading
MUSE-Wide cubes.

The individual field subpage on the top again provides links to the
headers and to download the datacube, but more importantly to the
quality control (QC) pages for each field. These QC pages show
important numbers that characterize each field, such as the observing
dates/conditions, the PSF estimation, the determined effective noise,
the sky brightness, offset information and pseudo-broad-band MUSE
images for different {\it HST} bands. The individual field pages
lastly show a table of CANDELS/GUO photometric objects in that field
on the basis of the field mapping. The table is searchable and
sortable and shows the position, various magnitudes and possible link
to an emission-line source. Clicking on the individual CANDELS/GUO ID
opens up yet another subpage for that individual photometric object,
which we describe below. 

\subsection{MUSE-Wide catalogs}\label{web-catalogs}

Clicking on the ``Catalogs'' tab opens a drop-down menu in which one
can select either the photometric or emission line catalog tables. The
creation of these catalogs and their respective 1D spectra has been
discussed in the respective sections in this paper (Section
\ref{emdetect} and Section \ref{tdose}).

Clicking on the photometric catalog subtab opens up the ``photometric
catalog'' subpage. At the top are the links to the complete 9,205
object catalog and to the identification catalog of objects brighter
than 24th magnitude (see Section \ref{photid}). Below the links is a
searchable and sortable table for all photometric catalog objects,
along with clickable links to an individual object subpage, link to
the corresponding field webpage (see previous section) and to a
possible linked emission-line source subpage.

The individual object subpages first show an image of a smoothed
spectrum (with a 9 pixel or 11.25~\AA\ Hanning window, also known as an
inverse cosine bell\footnote{\url{https://docs.scipy.org/doc/numpy-1.14.0/reference/generated/numpy.hanning.html}}),
a link to download a TDOSE-generated 1D spectrum and a link to download
a cutout $6\arcsec \times 6\arcsec$ minicube on the position of the
object generated on-the-fly. Lastly the object subpage shows $6\arcsec
\times 6\arcsec$ postage stamps of GOODS-S and CANDELS {\it HST}
images as well as a MUSE whitelight image centered on the objects
position. At the center of the postage stamp is a red circle with a
$1\arcsec$ radius to help identify the center of the postage stamp.

Clicking on the emission line catalog subtab then opens the ``Emission
line catalog'' subpage, which has the same layout as the photometric
catalog page, only that at the top the linked catalogs are the main
and emission line catalogs. The sortable and searchable table contains
the information for emission lines, again including links to either
emission-line objects subpages, the link to the field page and links
to the associated CANDELS-GUO subpage. The layout for the
emission-line object subpage is also very similar to the individual
photometric object subpages with images of the spectrum, postages,
links to the 1D spectrum and links to download a $6\arcsec \times
6\arcsec$ minicube are all present in this subpage. There are links to
both PSF-weighted and aperture 1D spectra centered on the LSDCat first
moment position.  

\subsection{Mini Cube cutout tool}

The ``Cut-Out'' subpage provides the user the opportunity to download
a custom made 3D minicube on positions centered on the MUSE-Wide
footprint. It will download only cube-data from one field according to
the Field mapping described in Section \ref{mapping} and includes all
extensions of the MUSE-Wide data cube (flux, effective noise, exposure
and whitelight). The minicube can have spatial sizes between 1\arcsec
and 20\arcsec and may encompass either a portion or the whole
wavelength range from 4750 -- 9350~\AA. The default is a
$6\arcsec\times6\arcsec$ using the whole wavelength range. The default
minicubes have file sizes of approximately 125MB. 
 
\subsection{SQL Query page and Database descriptors}\label{webdata}

The SQL query page allows the user to initiate more complicated search
queries using the various catalogs and images of MUSE-Wide. It also
allows cross-correlation to Simbad and VizieR. A simple SQL example on
how to select the 10 LAEs with the highest \Lya\, flux and their
respective F775W {\it HST} magnitudes with MUSE-Wide DR1 is shown
below: 

\lstset{language=SQL,
  basicstyle=\ttfamily,
  keywordstyle=\color{blue}\ttfamily,
  stringstyle=\color{red}\ttfamily,
  commentstyle=\color{green}\ttfamily,
  backgroundcolor = \color{light-gray}}
\begin{lstlisting}
SELECT m.unique_id, m.guo_id, m.ra, m.dec, 
m.z, lya.f_3kron AS flux_lya, 
phot.f775w_mag AS mag_775w 
FROM musewide_dr1.mw_44fields_main_table 
AS m 
JOIN musewide_dr1.mw_44fields_emline_table 
AS lya ON m.unique_id = lya.unique_id 
LEFT JOIN musewide_dr1.photometric_catalog 
as phot on m.guo_id = phot.guo_id 
WHERE lya.ident = 'Lya' 
ORDER BY flux_lya DESC 
LIMIT 10
\end{lstlisting}

The ``Database'' subpage provides the formal description of all the
data released in MUSE-Wide DR1. We opted to forgo a detailed description
of the tables in this paper, since it can be found in this tab. It is
possible to query for all entries presented within the various
released tables, catalogs and metadata. For both the emission-line
catalogs and the photometric catalogs a formal description each column
and data format of each catalog and the metadata linking the catalogs
and fields to each other can be found in this section. 

\end{document}